\documentclass[a4paper,10pt]{article}

  \usepackage{amsmath,amssymb,amsthm}

  \usepackage{graphicx}
  \graphicspath{ {./}{./fig/}{./fig/fig-01/}{./fig/fig-02/}{./fig/fig-03/}{./fig/fig-06/} }

  \usepackage{caption}
  \usepackage{subcaption}

  \usepackage{booktabs,tabularx}

  \usepackage{tikz,pgfplots}
  \usetikzlibrary{plotmarks}
  \usetikzlibrary{arrows}
  \usetikzlibrary{calc}
  \usetikzlibrary{decorations.text}

  \usetikzlibrary{external}
  \tikzset{external/system call={lualatex \tikzexternalcheckshellescape -halt-on-error -interaction=batchmode -jobname "\image" "\texsource"}}

  \usepackage{amsmath,amsfonts,amssymb,amsthm,mathrsfs}
  \usepackage{amsfonts,amssymb,amsthm,mathrsfs}

  \newcommand{\T}{^\mathsf{T}}                  
  \newcommand{\R}{\mathbb{R}}                   
  \newcommand{\cov}{\mathrm{cov}}               
  \providecommand{\norm}[1]{\|#1\|}             
  \newcommand{\diag}[1]{\mathrm{diag}(#1)}      
  \renewcommand{\vec}[1]{\mathrm{vec}(#1)}      
  \newcommand{\N}{\mathcal{N}}                  
  \newcommand{\GP}{\mathcal{GP}}                
  \newcommand{\kron}{\raisebox{1pt}{\ensuremath{\:\otimes\:}}} 
  \newcommand{\hadamard}{\circ}

  \usepackage{bm}
  
  \newcommand{\mbf}[1]{\mathbf{#1}}
  \newcommand{\vect}[1]{\mbf{#1}}
  \newcommand{\vectb}[1]{\bm{#1}}
  \newcommand{\mat}[1]{\mbf{#1}}
  \newcommand{\matb}[1]{\bm{#1}}

  \newtheorem{theorem}{Theorem}[section]
  \newtheorem{definition}[theorem]{Definition}

  \newtheorem{remark}[theorem]{Remark}

  \newcommand{\eg}{\textit{e.g.}}

  \usepackage{xcolor}

  \usepackage{graphicx}
  \graphicspath{ {./}{./fig/} }

  \usepackage{caption}
  \usepackage{subcaption}

  \usepackage{tikz,pgfplots}
  \usetikzlibrary{plotmarks}
  \pgfplotsset{compat=newest} 

  \usepackage{hyperref}
  \hypersetup{
    bookmarks=true,         
    pdfstartview={FitH},    
    colorlinks=true,        
    linkcolor=black,        
    citecolor=black,        
    filecolor=black,        
    urlcolor=black          
  }

  \usepackage{url}

  \usepackage{natbib}

  \usepackage[margin=1.5in]{geometry}

  \usepackage{authblk}
  \renewcommand\Affilfont{\itshape\small}
  \makeatletter
  \renewcommand\AB@affilsepx{;\quad \protect\Affilfont}
  \makeatother

  \title{Regularizing Solutions to the MEG Inverse Problem Using Space--Time Separable Covariance Functions}

  \author[a,b]{Arno Solin}

  \author[c]{Pasi Jyl{\"a}nki}

  \author[a,d]{Jaakko Kauram{\"a}ki}

  \author[e]{Tom Heskes}

  \author[c]{\\Marcel A. J. van Gerven}

  \author[f]{Simo S{\"a}rkk{\"a}}

  \affil[a]{Department of Neuroscience and Biomedical Engineering, 
    Aalto University, Espoo, Finland}

  \affil[b]{Department of Computer Science, 
    Aalto University, Espoo, Finland}

  \affil[f]{Department of Electrical Engineering and Automation, 
    Aalto University, Espoo, Finland}

  \affil[d]{Advanced Magnetic Imaging Centre, 
    Aalto University, Espoo, Finland}

  \affil[e]{Radboud University, Institute for Computing and Information Sciences, 
    Nijmegen, The Netherlands}

  \affil[c]{Radboud University, Donders Institute for Brain, Cognition and Behaviour, 
     Nijmegen, The Netherlands}

  \date{}

\begin{document}

\maketitle

\begin{abstract}
  In magnetoencephalography (MEG) the conventional approach to source reconstruction is to solve the underdetermined inverse problem independently over time and space. Here we present how the conventional approach can be extended by regularizing the solution in space and time by a Gaussian process (Gaussian random field) model. Assuming a separable covariance function in space and time, the computational complexity of the proposed model becomes (without any further assumptions or restrictions) $\mathcal{O}(t^3 + n^3 + m^2n)$, where $t$ is the number of time steps, $m$ is the number of sources, and $n$ is the number of sensors. We apply the method to both simulated and empirical data, and demonstrate the efficiency and generality of our Bayesian source reconstruction approach which subsumes various classical approaches in the literature.
\end{abstract}

\section{Introduction} \label{sec:intro}
Magnetoencephalography (MEG) is a non-invasive electrophysiological recording technique that is widely used in the neuroscience community~\citep{Cohen1972}. It provides data at high temporal resolution and the signal measured at each MEG sensor provides an instantaneous mixture of active neuronal sources.

Reconstruction of neuronal source time-series from sensor-level time-series requires the inversion of a generative model. This inverse problem is fundamentally ill-posed and additional assumptions with respect to the parameters of the neuronal sources are required. Two types of generative models have been proposed for neuromagnetic inverse problems: equivalent current dipole models and distributed source models \citep[see, \eg,][]{Baillet+Mosher+Leahy:2001}. Equivalent current dipole models can be estimated via Bayesian particle filtering methods \citep[see, \eg,][and the references therein]{ Sorrentino+Parkkonen+Pascarella+Campi+Piana:2009, Chen+Sarkka+Godsill:2015}. In the following, we focus on the distributed source models, which assume an extensive set of predefined source locations associated with unknown current amplitudes.

Over the past decades, various distributed inverse modelling techniques have been proposed and applied to neuroscience data, all of which impose their own assumptions. Commonly used techniques include the minimum-norm estimate (MNE) \citep{Hamalainen:1994} and the minimum-current estimate \citep{Uutela1999}, which both penalize source currents with large amplitudes but where the latter produces more sparse estimates, the low resolution electromagnetic tomography (LORETA) \citep{Pascual-Marqui:1994}, which promotes spatially smooth solutions, as well as different spatial scanning and beamforming techniques \citep{Veen1997,Mosher:1998}. Various extensions and modifications of these methods have been published including depth-bias compensations and standardized estimates \citep{Fuchs:1999, Dale:2000, Pascual-Marqui2002} and Bayesian frameworks for source and hyperparameter inference as well as model selection \citep{Trujillo:2004, Serinagaoglu:2005}. 

To improve localization accuracy, various different spatial priors have been proposed including spatially informed basis-functions whose parameters are learned from data in a Bayesian manner \citep{Phillips:2005,Mattout:2006}, and smoothness-promoting spherical harmonic and spline basis functions \citep{Petrov:2012} as well as various sparsity-promoting priors \citep{Gorodnitsky:1995, Sato:2004, Auranen:2005, Nummenmaa:2007, Haufe:2008, Wipf:2009, Wipf:2010, Lucka:2012, Babadi:2014}. Other approaches to improve the localization accuracy include combined models for simultaneous noise suppression \citep{Zumer:2007}, and models for potential inaccuracies in the forward model \citep{Stahlhut:2011}.

Most of the previously mentioned distributed source reconstruction techniques assume temporal independence between the source amplitudes, which does not make use of all the available prior information on the temporal characteristics of the neuronal currents. For this reason, many spatio-temporal techniques have been proposed including models based on different spatio-temporal basis functions and carefully designed or empirically learned Gaussian spatio-temporal priors \citep{Baillet:1997, Darvas:2001, Greensite:2003, Daunizeau:2006, Friston+Harrison+Daunizeau+Kiebel+Phillips+Trujillo-Barreto+Henson+Flandin+Mattout:2008, Litvak+Friston:2008, Trujillo:2008, Bolstad:2009, Limpiti:2009, Dannhauer:2013}. Many approaches are specifically targeted to obtain spatially sparse but temporally informed, smooth or oscillatory source estimates \citep{Cotter2005, Valdes:2009, Ou:2009, Haufe:2011, Zhang:2011, Tian:2012, Gramfort:2012, Gramfort:2013, Lina:2014, Castano:2015}. Dynamic extensions have been proposed also for spatially decoupled beamforming techniques \citep{Bahramisharif:2012, Woolrich:2013}.

Dynamical state-space models are a very intuitive way of defining spatio-temporal source priors. Many of the proposed approaches typically assume a parametric spatial structure derived from spatial proximity or anatomical connectivity, and a low-order multivariate autoregressive model for the temporal dynamics \citep{Yamashita:2004, Galka+Yamashita+Ozaki+Biscay+Valds-Sosa:2004, Long:2011, Lamus:2012, Pirondini:2012, Liu:2015, Fukushima:2015}. FMRI-informed spatial priors have also been proposed in \citep{Daunizeau:2007, Ou:2010, Henson:2010}.
Another class of spatio-temporal priors assumes that the sparsity is encoded in the the source scales rather  than amplitudes, which is the key assumption used by the multivariate Laplace prior \citep{Gerven2009} as well as the structured spatio-temporal spike-and-slab priors \citep{Andersen:2014, Andersen:2015}. 

In the present work, we show that high-quality Bayesian inverse solutions that employ various forms of spatio-temporal regularization can be obtained in an efficient manner by employing Gaussian processes (GPs) that use space--time separable covariance functions. These models are also referred to as Gaussian random fields. As infinite parametric models, GPs form a flexible class of priors for functions \citep[see, \eg,][]{Rasmussen+Williams:2006} and they have been successfully applied in various inverse problems \citep[see, \eg,][]{Tarantola:2004}. They are particularly suitable for spatio-temporal regularization because stationary GPs can also be interpreted as linear stochastic dynamical systems with a Gaussian state-space representation, which is the typical way of modelling physical systems \citep{Sarkka+Solin+Hartikainen:2013}. The inference can be done either using batch solutions or Kalman filtering and smoothing, and different GP priors can be easily combined to form, for example, temporally smooth and/or oscillatory sources with spatially smooth structure. 

Our approach is most closely related to previous approaches based on spatio-temporal basis functions or empirically informed prior covariance structures proposed by \citet{ Darvas:2001, Greensite:2003, Friston+Harrison+Daunizeau+Kiebel+Phillips+Trujillo-Barreto+Henson+Flandin+Mattout:2008, Trujillo:2008, Bolstad:2009, Limpiti:2009, Dannhauer:2013}. A key difference is that a finite set of basis functions is replaced by an infinite-dimensional process, or a combination of additive processes. The spatio-temporal characteristics of the resulting prior can be tuned by adapting covariance function hyperparameters in a data-driven empirical Bayesian manner \citep{Rasmussen+Williams:2006}.

Using both simulation studies and empirical data \citep{Kauramaki+Jaaskelainen+Hanninen+Auranen+Nummenmaa+Lampinen+Sams:2012}, we show that our approach is able to recover source estimates at a high degree of accuracy. Moreover, we show that various existing source reconstruction approaches \citep[\eg, ][]{Hamalainen:1994, Petrov:2012} are special cases of our framework that use particular spatial covariance functions. Various methodological developments ensure that our approach scales well with the number of sources and sensors. Furthermore, since the framework implements a Bayesian approach we also gain access to uncertainty estimates about the location of the recovered sources, and we can determine maximum posteriori estimates of the model hyperparameters in a principled manner.

Summarizing, as demonstrated on simulated as well as empirical data, the generality and efficiency of our framework open up new avenues in solving MEG inverse problems.

\section{Methods and Materials}
\subsection{MEG forward problem}
In magnetoencephalography (MEG) the biomagnetic fields produced by weak currents flowing in neurons are measured non-invasively with multichannel gradio- and magnetometer systems. For reviews on MEG methodology see, for example, \citet{Hamalainen+Hari+Ilmoniemi+Knuutila+Lounasmaa:1993} and \citet{Baillet+Mosher+Leahy:2001}. For now, we disregard any further information on how the model is described and derived using physical constraints, and we only consider a general-form linear forward model. Standard references on the mathematical model for MEG are \citet{Sarvas:1987} and \citet{Geselowitz:1970}.

Following the notation of \citet{Sarvas:1987}, we consider a forward model with a number of $n_\mathrm{m}$ sources (size of the brain mesh) and $n_\mathrm{n}$ sensors. Let the forward model with a number of $n_\mathrm{m}$ sources and $n_\mathrm{n}$ sensors be 
\begin{equation}
  \vect{b}_i = \mat{G} \vect{j}_i + \vectb{\varepsilon}_i,
\end{equation}
where $i$ indexes time, $\mat{G} \in \R^{n_\mathrm{n} \times n_\mathrm{m}}$ is the lead-field matrix, $\vect{b}_i \in \R^{n_\mathrm{n}}$ are the sensor readings, and $\vect{j}_i \in \R^{n_\mathrm{m}}$ the source values. The observations are corrupted by a noise term, that is assumed Gaussian, $\vectb{\varepsilon}_i \sim \N(\vect{0},\matb{\Sigma}_\mathrm{x})$. We assume the noise covariance matrix $\matb{\Sigma}_\mathrm{x}$ to be known, that is, not part of the estimation procedure.

As usual, only the quasi-static forward problem is considered in this context, and if we assume the forward problem to be time-invariant, we can rewrite the forward model for the entire range of observations in time. Let $\vect{b}_i$, $\vect{j}_i$, and $\vectb{\varepsilon}_i$ be the values at time instant $t_i, i=1,2,\ldots,n_\mathrm{t}$. This model can still be formulated by stacking the values into columns, such that
\begin{equation} \label{eq:meg-fwd}
  \mat{B} = \mat{G} \mat{J} + \mat{E},
\end{equation}
where $\mat{B} \in \R^{n_\mathrm{n} \times n_\mathrm{t}}$ corresponds to the sensor reading columns, and $\mat{J} \in \R^{n_\mathrm{m} \times n_\mathrm{t}}$ to the column-wise source values. The noise term $\mat{E}$ consists of columns, $\vectb{\varepsilon}_i \sim \N(\vect{0},\matb{\Sigma}_\mathrm{x})$, for $i = 1,2,\ldots,n_\mathrm{t}$. Equivalently we can write the observation model \eqref{eq:meg-fwd} in a vectorized form as
\begin{equation} \label{eq:meg-fwd-vec}
  \vec{\mat{B}} = \vec{\mat{G} \mat{J}} + \vec{\mat{E}} = ( \mat{I}_{n_\mathrm{t}} \kron \mat{G}) \vec{\mat{J}} + \vec{\mat{E}},
\end{equation}
where $\vec{\mat{E}} \sim \mathcal{N} (\vect{0}_{n_\mathrm{t} n_\mathrm{n} \times 1}, \mat{I}_{n_\mathrm{t}} \kron \matb{\Sigma}_\mathrm{x})$. Operator $\vec{\mat{B}}$ denotes the vectorization of matrix $\mat{B}$, which is obtained by stacking the columns of $\mat{B}$ on top of one another and results in a $n_\mathrm{t} n_\mathrm{n}$ column vector, and $\mat{I}_{n_\mathrm{t}} \kron \mat{G}$ denotes the Kronecker product, which is obtained by stacking together submatrices $[\mat{I}_{n_\mathrm{t}}]_{i,j} \mat{G}$ vertically for $i=1,2,\ldots,n_\mathrm{t}$ and horizontally for $j=1,2,\ldots,n_\mathrm{t}$ \citep[see, \eg,][]{Horn+Johnson:2012}.

Without loss of generality, we consider a model that has been whitened by the eigenvalue decomposition of the noise covariance matrix: $\matb{\Sigma}_\mathrm{x} = \mat{U}_\mathrm{x} \mat{D}_\mathrm{x} \mat{U}_\mathrm{x}\T$, 
where we retain only the positive eigenvalues. Multiplying \eqref{eq:meg-fwd} from the left by $\matb{\Lambda}_\mathrm{x}^{-1/2} \mat{U}_\mathrm{x}\T$ gives a whitened model 
$\tilde{\mat{B}} = \tilde{\mat{G}} \mat{J} + \tilde{\mat{E}}$ 
such that 
$\tilde{\mat{B}} = \mat{D}_\mathrm{x}^{-1/2} \mat{U}_\mathrm{x}\T \mat{B}$,
$\tilde{\mat{G}} = \mat{D}_\mathrm{x}^{-1/2} \mat{U}_\mathrm{x}\T \mat{G}$, 
and $\vec{\tilde{\mat{E}}} \sim \mathcal{N}(\vect{0}, \mat{I}_{n_\mathrm{m} n_\mathrm{t}} )$. 
For the remainder of the paper we will assume that the model is whitened and drop the tildes to simplify the notation.

It is possible to extend the proposed approach to include an estimate of the temporal structure of the noise in the form of a $n_\mathrm{t} \times n_\mathrm{t}$ covariance matrix $\matb{\Sigma}_\mathrm{t}$. In this case we write the noise covariance of the vectorized model \eqref{eq:meg-fwd-vec} as a Kronecker product of a spatial and a temporal covariance matrix, that is, $\cov(\vec{\mat{E}}) = \matb{\Sigma}_\mathrm{t} \kron \matb{\Sigma}_\mathrm{x}$. 
Computing the eigendecomposition $\matb{\Sigma}_\mathrm{t} = \mat{U}_\mathrm{t} \mat{D}_\mathrm{t} \mat{U}_\mathrm{t}\T$ and multiplying the vectorized model \eqref{eq:meg-fwd-vec} from the right by $\mat{D}_\mathrm{t}^{-1/2} \mat{U}_\mathrm{t}\T \kron \mat{D}_\mathrm{x}^{-1/2} \mat{U}_\mathrm{x}\T$, results in a whitened model $\vec{\tilde{\mat{B}}} = ( \mat{D}_\mathrm{t}^{-1/2} \mat{U}_\mathrm{t}\T \kron \tilde{\mat{G}}) \vec{\mat{J}} +\vec{\tilde{\mat{E}}}$, where $\tilde{\mat{B}} = \mat{D}_\mathrm{x}^{-1/2} \mat{U}_\mathrm{x}\T \mat{B} \mat{U}_\mathrm{t} \mat{D}_\mathrm{t}^{-1/2}$. All the subsequent results can be derived using this modified model with the main requirement being that the observation matrix is of spatio-temporal Kronecker form.

\subsection{Gaussian process based regularization in space and time}
Our interest is in the \emph{inverse problem} of finding the state of the sources given the sensor readings. The number of sources $n_\mathrm{m}$ (typically in thousands) is much larger than the number of sensors $n_\mathrm{n}$ (typically some hundreds), making the inverse problem underdetermined. Thus, in general, its solution is non-unique. Therefore we will regularize the solution by injecting prior knowledge concerning properties of the unknown underlying state of the neuronal generators of the MEG signal in the cerebral cortex.

We employ Gaussian process (GP) priors in order to regularize the solution of the inverse problem in space and time. In this section we first give the general form of the method, and then we show how the problem can be reformulated under the assumption of separable covariance functions. Separability in this context means the covariance structure factors into a purely spatial and purely temporal component. This does not restrain the model from coupling spatial and temporal effects in source-space, but restrains the model from enforcing coupling between arbitrary spatial and temporal locations explicitly.

The prior assumptions of the latent neuronal behavior on the cortex are encoded by a  Gaussian process prior (Gaussian random field), which has structure both in space and time. We define the Gaussian process \citep[see, \eg,][]{Rasmussen+Williams:2006, Tarantola:2004} as follows.

\begin{definition}[Gaussian process]
The process
    $f(\vect{r}) \sim \GP(\mu(\vect{r}), \kappa(\vect{r},\vect{r}'))$,
is a Gaussian process on $\mathcal{X}$ with a mean function $\mu: \mathcal{X} \to \R$, and a covariance function (kernel) $\kappa: \mathcal{X}\times\mathcal{X} \to \R$, if any finite collection of function values has a joint multivariate Gaussian distribution such that $(f(\vect{r}_1),f(\vect{r}_2),\ldots,f(\vect{r}_\mathrm{n}))\T \sim \N(\vectb{\mu},\mat{K})$, where $\mat{K}_{i,j} = \kappa(\vect{r}_i,\vect{r}_j)$ and $\vectb{\mu}_i = \mu(\vect{r}_i)$, for $i,j=1,2,\ldots,n$. \sloppy
\end{definition}

Consider the inputs consisting of both the spatial and temporal points on the cerebral cortex, $\vect{r} = (\vect{x}, t)$. We assign the following zero-mean Gaussian process prior (without loss of generality)
\begin{equation}
  f(\vect{x},t) \sim \GP(0,\kappa(\vect{x}, t; \vect{x}', t'))
\end{equation}
such that $\mat{J}_{j,i} = f(\vect{x}_j,t_i)$ and assume that the observations are given by the forward model as defined in the previous section:
\begin{equation}
  \vect{b}_i = \mat{G} 
    \big( f(\vect{x}_1,t_i) ~ f(\vect{x}_2,t_i) ~ \cdots ~ f(\vect{x}_{n_\mathrm{m}},t_i) \big)\T
  + \vectb{\varepsilon}_i.
\end{equation}
We can interpret the problem as a Gaussian linear model, with a marginal Gaussian distribution for $\mat{J}$ and a conditional Gaussian distribution for $\mat{B}$ given $\mat{J}$ of the form 
\begin{align*}
  \vec{\mat{J}} &\sim \N(\vect{0}, \mat{K}), \\
  \vec{\mat{B}} \mid \vec{\mat{J}} &\sim \N(\vec{\mat{G}\mat{J}}, \mat{I}_{n_\mathrm{m}}), 
\end{align*}
where the elements of the prior covariance matrix are given by $\mat{K}_{(i-1)n_\mathrm{m}+j, (k-1)n_\mathrm{m}+l} = \kappa(\vect{x}_j,t_i; \vect{x}_l, t_k)$.\sloppy

Following the above model, we can analytically compute the posterior distribution of $\mat{J}$ by conditioning on $\mat{B}$. Under the functional space view of the GP regression problem, we consider an infinite set of possible function values and predict $f(\vect{x}_*,t_*)$ for any $(\vect{x}_*,t_*)$ by integrating over $\mat{J}$ and conditioning on $\mat{B}$. The solution to the Gaussian process regression prediction problem can be given as follows. Consider an unknown latent source reading $f(\vect{x}_*,t_*)$ at $(\vect{x}_*,t_*)$. Conditioning on the observations $\mat{B}$ gives predictions $f(\vect{x}_*,t_*) \sim \N(\mathbb{E}[f(\vect{x}_*,t_*)], \mathbb{V}[f(\vect{x}_*,t_*)])$, where the mean and variance are given by
\begin{align}
  \mathbb{E}[f(\vect{x}_*,t_*)] &= 
    \vect{k}_*\T \mat{H}\T 
    (\mat{H} \mat{K} \mat{H}\T 
     + \mat{I})^{-1} \vec{\mat{B}},      \label{eq:prediction1}\\
  \mathbb{V}[f(\vect{x}_*,t_*)] &= 
    \kappa(\vect{x}_*,t_*; \vect{x}_*,t_*) 
    - \vect{k}_*\T \mat{H}\T 
    (\mat{H} \mat{K} \mat{H}\T 
     + \mat{I})^{-1} \mat{H} \vect{k}_*,     \label{eq:prediction2}
\end{align}
where $\vect{k}_*$ is an $n_\mathrm{m}n_\mathrm{t}$-dimensional vector with the $(i-1)n_\mathrm{m} +j$:th entry being $\kappa(\vect{x}_*, t_*;\vect{x}_j, t_i)$ and $\mat{H} = \textrm{blkdiag}(\mat{G},\ldots,\mat{G}) = \mat{I}_{n_\mathrm{t}} \kron \mat{G}$ is an $n_\mathrm{n}n_\mathrm{t} \times n_\mathrm{m}n_\mathrm{t}$ matrix. This result is readily available. These equations are derived, for example, in \citep[p.~93]{Bishop:2006}.

\subsection{Spatio-temporal regularization of space-time separable GPs}
For implementing Equations \eqref{eq:prediction1} and \eqref{eq:prediction2}, the dimensionality of the problem becomes an issue. For example, consider an MEG dataset of moderate size with $n_\mathrm{n} = 300$ and $n_\mathrm{t} = 1000$ observations. This means that the $n_\mathrm{n}n_\mathrm{t} \times n_\mathrm{n}n_\mathrm{t}$ matrix $(\mat{H} \mat{K} \mat{H}\T 
     + \mat{I})^{-1}$ to invert would be of size $300,\!000 \times 300,\!000$, which renders its practical use largely impossible. 

To circumvent this problem, in the following, we assume that the covariance structure of the underlying GP is separable in space and time. In the prewhitening step, we required also similar separability for the noise covariance. The following separable structure of $\kappa(\cdot,\cdot)$ is assumed:
\begin{equation} \label{eq:gp_cf_sep}
  \kappa(\vect{x}, t; \vect{x}', t') 
    = \kappa_\mathrm{x}(\vect{x}, \vect{x}') \, \kappa_\mathrm{t}(t, t'),
\end{equation}
meaning that the kernel is separable in space and time. The covariance matrix corresponding to this covariance function is thus
\begin{equation}
  \mat{K} = \mat{K}_\mathrm{t} \kron \mat{K}_\mathrm{x},
\end{equation}
where the temporal and spatial covariance matrices are defined as $[\mat{K}_\mathrm{t}]_{i,j} = \kappa_\mathrm{t}(t_i,t_j)$, for $i,j=1,2,\ldots,n_\mathrm{t}$, and $[\mat{K}_\mathrm{x}]_{i,j} = \kappa_\mathrm{x}(\vect{x}_i,\vect{x}_j)$, for $i,j=1,2,\ldots,n_\mathrm{m}$. We write down the eigendecompositions of the following covariance matrices: $\mat{K}_\mathrm{t} = \mat{V}_\mathrm{t} \matb{\Lambda}^\mathrm{t} \mat{V}_\mathrm{t}\T$ and $\mat{G}\mat{K}_\mathrm{x}\mat{G}\T = \mat{V}_\mathrm{x} \matb{\Lambda}^\mathrm{x} \mat{V}_\mathrm{x}\T$. We denote the eigenvalues of the matrices $\matb{\Lambda}^\mathrm{t}_{i,i} = \lambda^\mathrm{t}_i$ and $\matb{\Lambda}^\mathrm{x}_{j,j} = \lambda^\mathrm{x}_j$.
This brings us to the key theorem of this paper, a proof of which can be found in~\ref{appendixA}.

\begin{theorem}[Spatio-temporal regularization] \label{the:prediction}
  Consider an unknown latent source reading $f(\vect{x}_*, t_*)$ in a source mesh point $\vect{x}_*$ at an arbitrary time-instance $t_*$. Conditioning on all the observations $\mat{B}$ and integrating over $\mat{J}$ gives a predictive distribution 
  \begin{equation} \label{eq:posterior}
    f(\vect{x}_*, t_*) \sim \N\left(\mathbb{E}[f(\vect{x}_*, t_*)], \mathbb{V}[f(\vect{x}_*, t_*)] \right)
  \end{equation}
  where the mean and variance are given by
\begin{align}
  \mathbb{E}[f(\vect{x}_*,t_*)] &= 
    \vect{k}_{\mathrm{x}_*}\T\mat{G}\T\mat{V}_\mathrm{x} 
    (\matb{\Pi} \hadamard \mat{V}_\mathrm{x}\T\mat{B}\mat{V}_\mathrm{t})
    \mat{V}_\mathrm{t}\T \vect{k}_{\mathrm{t}_*}, \label{eq:mean} \\
  \mathbb{V}[f(\vect{x}_*,t_*)] &=
    \kappa_\mathrm{x}(\vect{x}_*,\vect{x}_*)\,\kappa_\mathrm{t}(t_*,t_*) 
       -[(\vect{k}_{\mathrm{x}_*}\T\mat{G}\T\mat{V}_\mathrm{x}) \hadamard
     (\vect{k}_{\mathrm{x}_*}\T\mat{G}\T\mat{V}_\mathrm{x})]
    \matb{\Pi}
    [(\mat{V}_\mathrm{t}\T \vect{k}_{\mathrm{t}_*}) \hadamard
     (\mat{V}_\mathrm{t}\T \vect{k}_{\mathrm{t}_*})], \label{eq:variance}
\end{align}
where $\vect{k}_{\mathrm{x}_*}$ is an $n_\mathrm{m}$-dimensional vector with the $j$th entry being $\kappa_\mathrm{x}(\vect{x}_*,\vect{x}_j)$, $\vect{k}_{\mathrm{t}_*}$ is an $n_\mathrm{t}$-dimensional vector with the $i$th entry being $\kappa_\mathrm{t}(t_*,t_i)$, and $\mat{B}$ is the matrix of  $n_\mathrm{n} \times n_\mathrm{t}$ measurements. The matrix $\mat{\Pi}$ has elements $\matb{\Pi}_{j,i} = {1}/{(\lambda^\mathrm{x}_j \lambda^\mathrm{t}_i + 1)}$, and by `$\hadamard\!$' we denote the Hadamard (element-wise) product.
\end{theorem}

Compared to traditional point-estimate based inverse solutions, Theorem~\ref{the:prediction} offers a flexible and computationally efficient Bayesian alternative, which enables principled probabilistic posterior summaries of the sources, efficient gradient-based estimation of regularization hyperparametes, and implementation of various existing regularizers as described in the following sections.

\subsection{Quantifying uncertainty about source estimates}
Probabilistic posterior summaries can be computed using the Gaussian marginal posterior \eqref{eq:posterior} for any chosen time instant and source location. For example, instead of summarizing only the mean $\mathbb{E}[\mat{J}_{j,i}]$, where $\mat{J}_{j,i} = f(\vect{x}_j, t_i)$, we can compute the marginal posterior probability for the source activation to be positive as $p(\mat{J}_{j,i} \ge 0 \mid \mat{B}) = \Phi( \mathbb{E}[\mat{J}_{j,i}]/ \mathbb{V}[\mat{J}_{j,i}] )$, where $\Phi(\cdot)$ is the Gaussian cumulative distribution function. This probability equals one if all the marginal probability density is located strictly at the positive real axis and zero in the opposite case.

\subsection{Encoding prior beliefs with covariance functions}
Within the GP framework, prior beliefs on source activations in space and time are encoded via different choices of kernels (covariance functions). In this section, we explore a number of alternative separable kernels functions of the form $\kappa(\vect{x},t; \vect{x}',t') = \kappa_\mathrm{x}(\vect{x}, \vect{x}') \kappa_\mathrm{t}(t, t')$. We first consider spatial kernels $\kappa_\mathrm{x}(\vect{x}, \vect{x}')$ and assume temporal independence such that the temporal kernel is defined as $\kappa_\mathrm{t}(t, t') = \delta(t-t')$, where $\delta(x) = 1$ if $x=0$ and zero otherwise. We will show that some of these spatial kernels correspond to (Bayesian variants of) source reconstruction algorithms that are used in the literature. After considering spatial kernels, we will consider an example of a spatio-temporal kernel in which we drop the temporal independence assumption.

\subsubsection{Minimum norm estimation}
If, next to temporal independence, we assume spatial independence, we obtain a covariance function that corresponds to a {\em minimum norm estimate} (MNE) solution~\citep{Hamalainen:1994}. This covariance function makes use of the kernel:
\begin{equation} \label{eq:cov_indep}
  \kappa_{\textrm{MNE}}(\vect{x}, \vect{x}') = \gamma^2\,\delta(\norm{\vect{x}-\vect{x}'}),
\end{equation}
where $\gamma^2$ is a magnitude parameter that controls the prior variance of the source current amplitudes. Coupled with the temporal independence assumption, this kernel results in diagonal covariance matrices $\mat{K}_\mathrm{x}= \gamma^2 \mat{I}_{n_\mathrm{m}}$ and $\mat{K}_\mathrm{t}=\mat{I}_{n_\mathrm{t}}$. 
\begin{remark}
  If we assume kernel \eqref{eq:cov_indep} and compute the predictions in Lemma~\ref{the:prediction} for all grid locations $i=1,2,\ldots,n_\mathrm{m}$ at time instant $t_i$, the eigendecompositions result in $\mat{V}_\mathrm{t}=\mat{I}_{n_\mathrm{t}}$, $\matb{\Lambda}_\mathrm{t}=\mat{I}_{n_\mathrm{t}}$ and
  $\gamma^2 \mat{G}\mat{G}\T = \mat{V}_\mathrm{x} \matb{\Lambda}_\mathrm{x} \mat{V}_\mathrm{x}\T$,
  and the mean expressions convert into the so-called \emph{minimum norm estimate}
\begin{equation}
  \vect{j}_i = \mat{G}\T(\mat{G}\mat{G}\T + \gamma^{-2} \mat{I}_{n_\mathrm{m}})^{-1}\vect{b}_i,
\end{equation}
where $\gamma^{-2}$ assumes the role of the traditional regularization parameter.
\end{remark}

This shows that the presented framework can be used to compute a Bayesian equivalent of the MNE estimate.
In contrast to the traditional view, the Bayesian version offers also probabilistic uncertainty estimates of the source amplitudes via \eqref{eq:variance} as well as a convenient proxy for determining the regularization parameter from the observed data using the marginal likelihood as described in Section~\ref{sec:hyper_opt}.

\subsubsection{Exponential covariance function}
As Gaussian processes are formulated as non-parametric models, covariance functions can be interpreted as spanning the solution with an infinite number of possible basis functions. A canonical example is the \emph{exponential covariance function} which is given by the kernel
\begin{equation}
  \kappa_\text{EXP}(\vect{x},\vect{x}') = \gamma^2 \, \exp(-\norm{\vect{x}-\vect{x}'}/\ell)
\end{equation}
where $\gamma^2$ is a magnitude hyperparameter and $\ell$ is the characteristic length scale. Exponential covariance functions encode an assumption of continuous but not necessarily differentiable sample trajectories. In one dimension such processes are also known as Ornstein--Uhlenbeck processes \citep[see, \eg,][]{Rasmussen+Williams:2006}.

\subsubsection{Mat\'ern covariance function}
An alternative to the exponential covariance function is the \emph{Mat\'ern covariance function} (with smoothness parameter $\nu = 3/2$), which is given by the kernel
\begin{equation}
  \kappa_\text{MAT}(\vect{x},\vect{x}') = \gamma^2 \, \bigg(1 + \frac{\sqrt{3}\,\norm{\vect{x}-\vect{x}'}}{\ell} \bigg) \exp\bigg(- \frac{\sqrt{3}\,\norm{\vect{x}-\vect{x}'}}{\ell} \bigg)\:.
\end{equation}
This kernel produces estimates which are more smooth than those produced by an exponential prior. In the one-dimensional case, the Mat\'ern covariance function encodes the assumption that the sample trajectories are continuous once-differentiable functions.

\subsubsection{Gaussian covariance function}
Another often used formulation is given by the {\em Gaussian covariance function}, which uses the \emph{radial basis function} (RBF) kernel
\begin{equation}
  \kappa_\text{RBF}(\vect{x},\vect{x}') = \gamma^2\, \exp\bigg(-\frac{\norm{\vect{x}-\vect{x}'}^2}{2\ell^2}\bigg)\:.
\end{equation}
This kernel encodes the assumption of infinitely smooth (infinitely differentiable) processes.

\subsubsection{Harmony covariance function}
The exponential, Mat\'ern and Gaussian kernels are examples of nondegenerate kernels that have an infinite number of eigenfunctions~\citep{Rasmussen+Williams:2006}. We can alternatively make use of degenerate kernels that can be described in terms of a finite number of eigenfunctions. 

We here consider a model which assumes that the inference is done on a spherical surface and which has been considered in the literature before. The {\em Harmony} (HRM) model by \citet{Petrov:2012} is based on a projection of the inference problem onto spherical harmonical functions. Higher frequencies are suppressed in this approach by down-weighting them \citep{Petrov:2012}. Recalling that Cartesian coordinates $\vect{x}$ on the spherical hull correspond to $(\theta,\phi)$ in spherical coordinates, the Harmony model can be re-written as a weighted set of standard spherical harmonic basis functions 
\begin{equation}
  \kappa_\text{HRM}(\vect{x},\vect{x}') = \gamma^2 \sum_{\ell=0}^{\ell_\mathrm{max}} \sum_{m=-\ell}^{\ell} \mathrm{Y}_\ell^m(\theta,\phi) \frac{1}{1+\ell^p} \mathrm{Y}_\ell^m(\theta',\phi'),
\end{equation}
where $n_\textrm{b} = (\ell_\mathrm{max}+1)^2$ is the total number of spherical harmonic basis functions (terms in the sums), $\ell$ denotes the harmonic index, $\gamma^2$ is the magnitude hyperparameter, and $p$ a spectral weight (length-scale) hyperparameter. The fraction term defines the spectrum of the model. The Laplace's spherical harmonic functions are defined as \citep[see, \eg,][]{Arfken+Weber:2001}:
\begin{equation} \label{eq:spherical-harmonics}
  \mathrm{Y}_\ell^m(\theta,\phi) = 
  \sqrt{\frac{(2\ell+1)(\ell + |m|)!}{4\pi \, (\ell + |m|)!}} \,
  \mathrm{P}_\ell^{|m|}(\cos \theta)
  \begin{cases}
    1, & m=0, \\
    \sqrt{2} \cos(m \phi), & m>0, \\
    \sqrt{2} \sin(m \phi), & m<0,
  \end{cases}
\end{equation}
where $\mathrm{P}_{\ell}^{m}(\cdot)$ denotes the associated Legendre polynomials. Spherical harmonics are smooth and defined over the entire sphere. They can thus induce spurious correlations if too few basis functions are included in the model, or the weighting of the higher-frequency components is inconsistent. However, they span a complete orthonormal basis for the solution. \citet{Petrov:2012} used $n_\textrm{b}=121$ harmonic basis functions and chose $p=0.9$ by minimizing a measure on the localization error in a set of simulated inverse problems.

\subsubsection{Spline covariance function}
To avoid spurious correlations in the Harmony model, one option is to make the basis functions local. The {\em spline} (SPL) model by \citet{Petrov:2012} considers evenly spaced splines on the spherical hull, whose weighted superposition gives the following kernel:
\begin{equation}
  \kappa_\text{SPL}(\vect{x},\vect{x}') = \gamma^2 \, \sum_{j=1}^{n_\textrm{b}} K(\vect{x},\hat{\vect{r}}_j) \, K(\vect{x}',\hat{\vect{r}}_j)\:.
\end{equation}
Here $K(\vect{x},\hat{\vect{r}}_j)$ denotes the Abel--Poisson kernel centered at $\hat{\vect{r}}_j$, which is given by \citet{Freeden+Schreiner:1995}:
\begin{equation}\label{eq:abel-poisson}
  K(\vect{x},\hat{\vect{r}}_j) = 
  \frac{1-h^2}{4\pi\,(1+h^2-2h\, \vect{x}\T \hat{\vect{r}}_j)^{3/2}},
\end{equation}
where $h$ is a length-scale hyperparameter. Following \citet{Petrov:2012}, we use $n_\textrm{b}=162$ spherical splines uniformly positioned on both cortical surfaces with $\hat{\vect{r}}_j$ located at the nodes of the second subdivision of icosahedron meshing. The SPL model captures small-scale phenomena well, but is somewhat sensitive to where the splines are placed.

Increasing the number of splines $n_\textrm{b}$ diminishes the effect of the spline placement related artefacts. Using the connection between the Poisson kernel in multipole expansions and the Cauchy distribution, there is a relationship between this spline kernel and a more general class of covariance functions. If the number of splines $n_\textrm{b} \to \infty$, this degenerate kernel approaches the following nondegenerate kernel (up to a scaling factor):
\begin{equation}
  \kappa_\text{RQ}(\vect{x},\vect{x}') = \gamma^2\, \Bigg(1+\frac{\norm{\vect{x}-\vect{x}'}^2}{2\alpha\ell^2}\Bigg)^{-\alpha}\,,
\end{equation}
where the decay parameter $\alpha = 3/2$. This is known as the rational quadratic (RQ) covariance function \citep[see][for a similar parametrization]{Rasmussen+Williams:2006}. By this connection, we may recover the characteristic length-scale relation between the spline kernel and the family of nondegenerate kernels we have presented.  \citet{Petrov:2012} set the spline scale parameter to $h=0.8$ by minimizing a measure on the localization error in a set of simulated inverse problems. This corresponds to a characteristic length-scale of $\ell \approx 0.262$.

\subsubsection{Spatio-temporal covariance function}
So far, we assumed temporal independence. However, next to using different formulations for the spatial kernel, we can also use different formulations for the temporal kernel that drop the temporal independence assumption. Here, we only consider one such formulation of a spatio-temporal kernel that is written in terms of spatial and temporal exponential kernels:
\begin{equation}
  \kappa_\text{ST}(\vect{x},t; \vect{x}',t') = \gamma^2\, \exp(-\norm{\vect{x}-\vect{x}'}/\ell_\text{x}) \, \exp(-|t-t'|/\ell_\text{t})\,.
  \label{eq:spatiotemporal}
\end{equation}
Because the exponential kernel presumes continuity but not necessarily differentiability, we regard it as a good general purpose kernel for settings where there is no exact information about the smoothness properties of underlying source activations. However, it should be stressed that we use the exponential spatio-temporal kernel mainly for the purpose of illustration as the generality of our framework allows for arbitrary choices of spatial and temporal kernels.

\subsection{Hyperparameter optimization} \label{sec:hyper_opt}
The hyperparameters $\vectb{\theta}$ of the covariance function were earlier suppressed in the notation for brevity. In a typical spatio-temporal GP prior setting following \eqref{eq:spatiotemporal} the hyperparameter vector consists of the magnitude parameter and the spatial and temporal length-scales: $\vectb{\theta} = \{ \gamma, \ell_\text{x}, \ell_\text{t} \}$. For finding the hyperparameters, we consider the marginal likelihood $p(\vec{\mat{B}} \mid \vectb{\theta})$ of the model. The negative log marginal likelihood for the spatio-temporal model is
\begin{align} \label{eq:ml}
  \mathcal{L}(\vectb{\theta}) 
    = \frac{1}{2} \sum_{i=1}^{n_\mathrm{t}} \sum_{j=1}^{n_\mathrm{n}} 
         \log (\lambda^\mathrm{x}_j \lambda^\mathrm{t}_i + 1) 
    + \frac{1}{2} \vec{\mat{V}_\mathrm{x}\T\mat{B}\mat{V}_\mathrm{t}}\T 
         \matb{\Lambda}^{-1} \vec{\mat{V}_\mathrm{x}\T\mat{B}\mat{V}_\mathrm{t}} 
    + \frac{n_\mathrm{n} n_\mathrm{t}}{2} \log(2\pi).
\end{align}
The eigenvectors $\{ \mat{V}_\mathrm{x}, \mat{V}_\mathrm{t} \}$ and the eigenvalues $\{ \lambda^\mathrm{x}_j, \lambda^\mathrm{t}_i \}$ depend on the hyperparameters $\vectb{\theta}$, and in general re-calculating the eigendecompositions are required if the parameters change. However, for the magnitude parameter $\gamma$ this is not required, as it can be seen as a scaling factor of the prior covariance $\mat{K} = \gamma^2 ( \mat{K}_\mathrm{t}(\ell_\text{t}) \kron \mat{K}_\mathrm{x}(\ell_\text{x}) ) = \gamma^2 \mat{K}_\mathrm{t}(\ell_\text{t}) \kron \mat{K}_\mathrm{x}(\ell_\mathrm{x})$. This means that if the length-scales are fixed, the eigendecompositions need to be computed only once, and during the subsequent optimization $\gamma$ can be added as a multiplier to either all $\lambda^\mathrm{x}_j$s or all $\lambda^\mathrm{t}_i$s. Similarly, if either of the length-scale variables is fixed, the corresponding eigendecomposition does not need to be recomputed.

\subsubsection{Choosing the spatial and temporal length-scales}
The strongly ill-posed inverse problem and the potentially multimodal marginal likelihood $p(\mat{B} \mid \vectb{\theta})$ make a purely data-driven estimation of the hyperparameters challenging. In parametric models, the scale of the modeled phenomena is often tuned by choosing an appropriate number of basis functions. Under the non-parametric setting, however, the characteristic length-scale hyperparameter tunes scale-variability of the estimate.

One option to facilitate the hyperparameter estimation is to inject prior information into the model by using an informative prior $p(\vectb{\theta})$ and determine the maximum a posteriori estimate of $\vectb{\theta}$. A prior to the temporal length-scale $\ell_{\textrm{t}}$ could be chosen so that the spectral bandwidth of the source amplitudes is at some reasonable region. For example, in the empirical experiments we chose a fixed value for the temporal length-scale, $\ell_{\textrm{t}}=50$~ms, which enables the model to explain well the most quickly-varying peaks of the averaged event-related potentials. The solution is not sensitive to the choice of the temporal length-scale, because optimizing the magnitude parameter $\gamma$ compensates for small changes in $\ell_{\textrm{t}}$. This can be verified by comparing the predictive distribution of the sensor reading with the observed data.

A prior for the spatial length-scale can be chosen by acknowledging the natural spatial cutoff frequency that arises from the finite number of MEG channels by the Nyquist limit. For $n_\mathrm{n}=300$ and a distance measure defined on the unit sphere, this corresponds to a minimal spatial length-scale of $\ell_\mathrm{x}=0.5$ in the case of an exponential kernel. Another viable approach to choose the spatial length-scale could be to simulate various kinds of source activation patterns and choose the spatial length-scale value that maximizes the localization accuracy in terms of some suitable measure as proposed by \citet{Petrov:2012}. As presented in the previous sections, the chosen parameters in \citet{Petrov:2012} correspond to a characteristic length-scale of roughly $\ell_\mathrm{x} \approx 0.262$, while the low-rank nature of Harmony favors large-scale smooth activations. 

To avoid making too strong spatial smoothness assumptions, we fixed the spatial length-scale to $\ell_\mathrm{x} = 0.1$ in the empirical comparisons. This compensates for the non-ideal distance measures, which are not truly isotropic but given along the spherical cortex model. Optimizing the magnitude parameter with fixed temporal and spatial length-scales allows the model to explain the observed data well and compensates for small uncertainty regarding this choice of fixed length-scales.

\subsection{Computational complexity}
\label{sec:computational-complexity}
The most critical computations for evaluating the predictions \eqref{eq:mean} and \eqref{eq:variance}, as well as the marginal likelihood \eqref{eq:ml}, are the eigendecompositions $\mat{G}\mat{K}_\mathrm{x}\mat{G}\T = \mat{V}_\mathrm{x} \matb{\Lambda}^\mathrm{x} \mat{V}_\mathrm{x}\T$ and $\mat{K}_\mathrm{t} = \mat{V}_\mathrm{t} \matb{\Lambda}^\mathrm{t} \mat{V}_\mathrm{t}\T$, which scale as $\mathcal{O}(n_\mathrm{n}^3)$ and $\mathcal{O}(n_\mathrm{t}^3)$ respectively. The savings are significant compared with the naive implementation of \eqref{eq:prediction1} and \eqref{eq:prediction2} which scales as $\mathcal{O}(n_\mathrm{t}^3 n_\mathrm{n}^3)$. The eigendecompositions need to be recomputed only once after adjusting the covariance function hyperparameters that control $\mat{K}_\mathrm{x}$ and $\mat{K}_\mathrm{t}$. For fixed prior covariances, the decompositions need to be computed only once.

In practice, the decomposition of $\mat{G}\mat{K}_\mathrm{x}\mat{G}\T$ is insignificant because typically $n_\mathrm{n} \approx 300$. The decomposition of $\mat{K}_\mathrm{t}$ can become problematic if $n_\mathrm{t}$ grows too large but for typical MEG studies with up to few thousand time points per trial the inverse analysis can be done in few seconds using a standard PC. With small to mediocre $n_\mathrm{t}$, the additional cost for spatio-temporal GP inverse analysis is negligible compared to the standard MNE estimate.

\subsection{Simulated data} 

In order to compare the behaviour of the various introduced priors in the spatial dimension, we made use of simulated data. Data were simulated by simultaneously activating two distinctive source patches. The sources were located on opposite cerebral hemispheres, and together they measured  roughly 9.2~cm$^2$ in size (measured along the cortical surface).
Dipole orientations were fixed to be orthogonal to the cortical surface. The strength of the source dipoles was chosen such that they replicated the sensor data from the experimental recordings ($40 \cdot 10^{-10}$~Volts), and Gaussian noise with a covariance structure from a pre-measurement empty-room batch of data was added to the simulated data sets. Typical running times for our simulations using independent, spatial, temporal and (separable) spatio-temporal priors are summarized in Table~\ref{tab:time}.

\begin{table}[!t]
  \caption{Typical running times (including whitening and hyperparameter optimization for $\gamma^2$) as observed on a standard desktop computer with calculations over 80 data sets.}
  \label{tab:time}
  \centering
  \begin{tabular}{l c c}
    \toprule
	Method & Mean running time [s] & Max running time [s] \\
    \midrule
    MNE (independent) & 13.26 & 17.44 \\
    GP (spatial) & 25.37 & 32.26 \\
    GP (temporal) & 30.97 & 35.87 \\
    GP (spatio-temporal) & 42.11 & 51.67 \\
    \bottomrule
  \end{tabular}
\end{table}

\subsection{Empirical data}
In order to empirically validate our approach we made use of the data described in \citet{Kauramaki+Jaaskelainen+Hanninen+Auranen+Nummenmaa+Lampinen+Sams:2012}. The goal of this study was to further our understanding of selective attention to task-relevant sounds whilst ignoring background noise. In the present paper, we use a subset of the data in order to study source reconstruction results under different GP regularisation schemes. We briefly describe the main characteristics of these data. 

A subset of 10 subjects (7~males) out of 14~subjects was used in our analyses as their cortical surface reconstructions from whole-head anatomical MRI images were readily available. All subjects signed a written informed consent before the study. The measurements conformed to the guidelines of the Declaration of Helsinki, and the research was approved by the ethical committee in the Hospital District of Helsinki and Uusimaa. The task of the subjects was to attend either to the auditory or visual stream, and to respond by lifting a finger whenever an attended target stimulus occurred. The task was alternated every two minutes with instructions on screen, with two auditory and two visual tasks in each 8-minute block. This was repeated eight times with short breaks in between.

In this study we consider the auditory condition only. Subjects were presented with 300~ms tones of 1000~Hz (`standard', 90\% of sounds) and 1020 Hz (`target', 10\% of sounds) frequencies. The mean interstimulus interval from onset to onset was 2 seconds (range 1800--2200~ms). The tones were played either on their own (`no masker' condition) or with a masker sound (7 different maskers). The continuous masker sounds (16-bit, 48-kHz) were created by filtering in the frequency domain 10-minute Gaussian white noise with symmetrical stopbands or notches around 1000~Hz. Seven different notch widths were used in order to manipulate the impact of masking ($\pm$500~Hz, $\pm$300~Hz, $\pm$200~Hz, $\pm$150~Hz, $\pm$100~Hz, $\pm$50~Hz, $\pm$0~Hz). As the perceptual SNR was adjusted so that the sounds were played at individual hearing threshold, this notch width manipulation resulted in clearly poorer performance in the task with narrower notches. That is, both standard and deviant tones were harder to detect from the background noise masker. Each experimental block was run with the same continuous masker sound with the blocks jointly spanning all used notches and the no-masker background sound. 

Magnetoencephalography (MEG) data was acquired using a 306-channel whole-head neuromagnetometer (Vectorview, Elekta Neuromag Oy, Helsinki, Finland). MEG data were recorded at 2000~Hz. To detect eye blinks and movements, one electro-oculogram (EOG) channel was recorded with the electrodes placed below and on the outer canthus of the left eye. Auditory-evoked responses to both standard and deviant tones were averaged. In the data analysis, whitened averaged $-$100--800~ms time-locked epochs were used. Epochs whose amplitudes exceeded predefined thresholds were excluded from further analysis. A prestimulus baseline of 200~ms was used to remove DC offset, and a 40 Hz lowpass filter was used during averaging. Three-dimensional locations of left and right preauricular points and nasion, four head-position indicator (HPI) coils, and a number of extra points from the scalp were used for estimating head position and alignment with cortical surface reconstruction obtained using FreeSurfer (v4.5.0, \url{http://surfer.nmr.mgh.harvard.edu}) software. A spherical head model was used to compute the lead-field matrix \citep{Hamalainen+Hari+Ilmoniemi+Knuutila+Lounasmaa:1993}.

\citet{Kauramaki+Jaaskelainen+Hanninen+Auranen+Nummenmaa+Lampinen+Sams:2012} showed that selective attention to sounds increased the M100 response amplitude and at the same time enhanced the sustained response. It also showed that M100 response peak amplitudes increased and latencies shortened with increasing notch width. The latter result will be used to compare source reconstruction approaches in the present study.

\section{Results}
In order to test our spatio-temporal GP framework for solving the MEG inverse problem, we compare how well the introduced kernels are able to recover sources of interest. To this end we make use of both simulations and empirical data.

\subsection{Simulation study} \label{sec:simulation-study}
We here test how the choice of spatial prior affects the spatial correlation structure of the reconstructed source configuration (assuming temporal independence). This prior information is controlled by the choice of spatial covariance function. As presented in the previous sections, the highly non-isotropic nature of the cerebral cortex can be warped through a non-linear coordinate transformation. We used a spherical inflation of the cerebral cortex for mapping the source distances.

\begin{figure*}[!t]
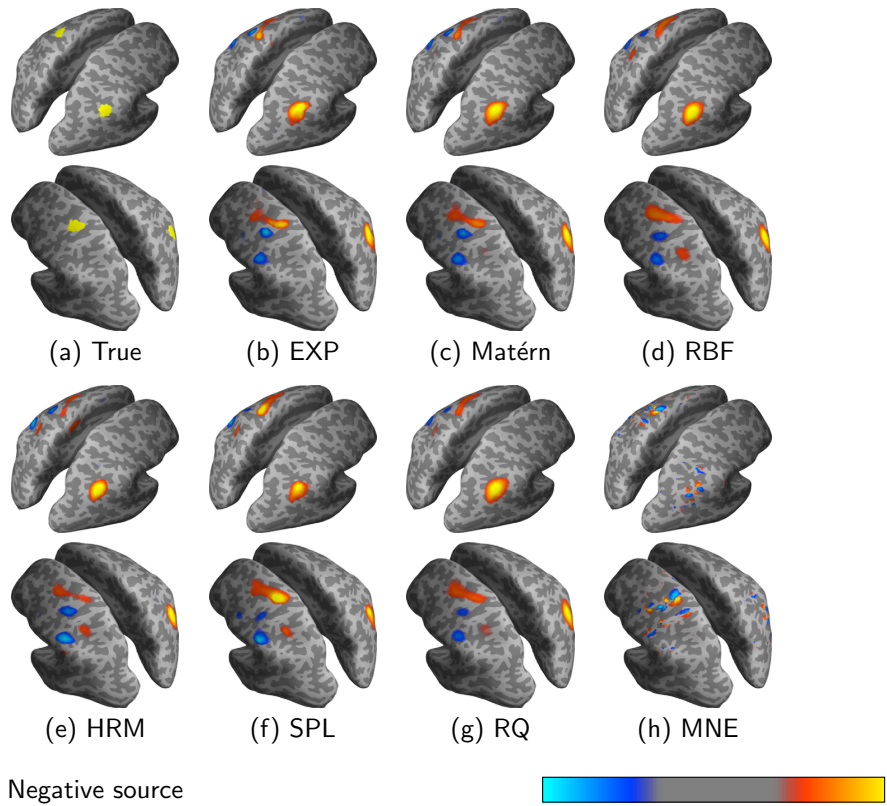


  \tikzsetnextfilename{figure-01}

  \centering\sffamily
  \begin{tikzpicture}

    \def\methodsa{mag-sim1-true,mag-sim1-gpkron-exp,mag-sim1-gpkron-matern32,mag-sim1-gpkron-sexp}
    \def\methodsb{mag-sim1-gpkron-harmony,mag-sim1-gpkron-spl,mag-sim1-gpkron-rq,mag-sim1-mne}
    \def\namesa{{(a)~True},{(b)~EXP},{(c)~Mat\'ern},{(d)~RBF}}
    \def\namesb{{(e)~HRM},{(f)~SPL},{(g)~RQ},{(h)~MNE}}
    \def\xspacing{2.6}

    \foreach[count=\i] \x in \methodsa {
      \node at ({\xspacing*\i},{0}) {\includegraphics[width=2.2cm]{fig/fig-01/\x-a.png}};
      \node at ({\xspacing*\i},{-2.2}) {\includegraphics[width=2.2cm]{fig/fig-01/\x-b.png}};
    }

    \foreach[count=\i] \x in \methodsb {
      \node at ({\xspacing*\i},{-5}) {\includegraphics[width=2.2cm]{fig/fig-01/\x-a.png}};
      \node at ({\xspacing*\i},{-7.2}) {\includegraphics[width=2.2cm]{fig/fig-01/\x-b.png}};
    }

    \foreach[count=\i] \x in \namesa {
      \node at ({\xspacing*\i},{-3.6}) {\x};
    }

    \foreach[count=\i] \x in \namesb {
      \node at ({\xspacing*\i},{-8.6}) {\x};
    }

    \node at (\xspacing,-9.4) {Negative source};
    \node at (4*\xspacing,-9.4) {Positive source};

    \node at (6.05,-9.5) {%
    \begin{axis}[
      hide axis,
      scale only axis,
      height=0cm,
      width=0cm,
      colormap={custom}{
      rgb255(0pt)=(0, 255, 255);
      rgb255(1pt)=(0, 243, 255);
      rgb255(2pt)=(0, 231, 255);
      rgb255(3pt)=(0, 219, 255);
      rgb255(4pt)=(0, 206, 255);
      rgb255(5pt)=(0, 194, 255);
      rgb255(6pt)=(0, 182, 255);
      rgb255(7pt)=(0, 170, 255);
      rgb255(8pt)=(0, 158, 255);
      rgb255(9pt)=(0, 146, 255);
      rgb255(10pt)=(0, 134, 255);
      rgb255(11pt)=(0, 121, 255);
      rgb255(12pt)=(0, 109, 255);
      rgb255(13pt)=(0, 97, 255);
      rgb255(14pt)=(0, 85, 255);
      rgb255(15pt)=(0, 73, 255);
      rgb255(16pt)=(2, 62, 253);
      rgb255(17pt)=(26, 64, 229);
      rgb255(18pt)=(50, 72, 205);
      rgb255(19pt)=(74, 84, 181);
      rgb255(20pt)=(98, 100, 157);
      rgb255(21pt)=(122, 122, 133);
      rgb255(22pt)=(128, 128, 128);
      rgb255(23pt)=(128, 128, 128);
      rgb255(24pt)=(128, 128, 128);
      rgb255(25pt)=(128, 128, 128);
      rgb255(26pt)=(128, 128, 128);
      rgb255(27pt)=(128, 128, 128);
      rgb255(28pt)=(128, 128, 128);
      rgb255(29pt)=(128, 128, 128);
      rgb255(30pt)=(128, 128, 128);
      rgb255(31pt)=(128, 128, 128);
      rgb255(32pt)=(128, 128, 128);
      rgb255(33pt)=(128, 128, 128);
      rgb255(34pt)=(128, 128, 128);
      rgb255(35pt)=(128, 128, 128);
      rgb255(36pt)=(128, 128, 128);
      rgb255(37pt)=(128, 128, 128);
      rgb255(38pt)=(128, 128, 128);
      rgb255(39pt)=(128, 128, 128);
      rgb255(40pt)=(128, 128, 128);
      rgb255(41pt)=(128, 128, 128);
      rgb255(42pt)=(128, 128, 128);
      rgb255(43pt)=(139, 116, 116);
      rgb255(44pt)=(163, 96, 92);
      rgb255(45pt)=(187, 80, 68);
      rgb255(46pt)=(211, 70, 44);
      rgb255(47pt)=(235, 63, 20);
      rgb255(48pt)=(255, 64, 0);
      rgb255(49pt)=(255, 76, 0);
      rgb255(50pt)=(255, 88, 0);
      rgb255(51pt)=(255, 100, 0);
      rgb255(52pt)=(255, 112, 0);
      rgb255(53pt)=(255, 124, 0);
      rgb255(54pt)=(255, 137, 0);
      rgb255(55pt)=(255, 149, 0);
      rgb255(56pt)=(255, 161, 0);
      rgb255(57pt)=(255, 173, 0);
      rgb255(58pt)=(255, 185, 0);
      rgb255(59pt)=(255, 197, 0);
      rgb255(60pt)=(255, 209, 0);
      rgb255(61pt)=(255, 222, 0);
      rgb255(62pt)=(255, 234, 0);
      rgb255(63pt)=(255, 246, 0);
    },
    point meta min=0,
    point meta max=1,
    colorbar horizontal,
    colorbar style={
        width=4.5cm,
        height=.33cm,
        xtick={0,1},
        xticklabels={},
        xtick style={draw=none}
    }]
    \addplot [draw=none] coordinates {(0,0)};
  \end{axis}%
  };

  \end{tikzpicture}

  \caption{Spatial source reconstructions for a simulated experiment with two source patches. The following spatial covariance structures were tested using the spherically inflated cortical distance measures: (a)~is the ground truth, (b)--(d) are the presented nondegenerate kernels, (e)--(f) correspond to the methods HRM and SPL by Petrov, (g)~is the rational quadratic covariance function, which is recovered in the limit of points in (f) going to infinity, and (h) is the MNE solution obtained with the independent model. Warm colors indicate positive sources, cold colors negative sources. An equal area of activations are shown for each method corresponding to the strongest values.}
  \label{fig:sim-spatial}
\end{figure*}

We compared the following spatial covariance structures: MNE (spatially independent prior), the exponential covariance function, the Mat\'ern covariance function, and the squared-exponential covariance function. For all methods, the regularization scale (magnitude) parameter was optimized by maximizing the marginal likelihood. The length-scales for HRM and SPL were chosen following the suggestions of \citet{Petrov:2012}, and the length-scale $\ell$ of the GP covariances was chosen so that it approximately matches the decrease-rate of the spectral density of HRM covariance. Figure~\ref{fig:sim-spatial} shows the ground truth and reconstructions of the source estimates using various covariance structures in the GP reconstruction. The reconstructed sources were thresholded such that the strongest $2.5\%$ of all sources are shown for each method. This puts the different GP prior covariance functions on the same line.

Results show that all of the chosen spatial priors lead to reasonable results, in the sense that the ground truth sources are properly localised. At the same time results show that the spatially independent MNE solution yields a non-smooth result where the sign of the reconstructed source tends to fluctuate quite randomly. In contrast, the spatial GP-regularised solutions are spatially smooth showing just a few other sources with spurious activation.

\begin{figure*}[!t]
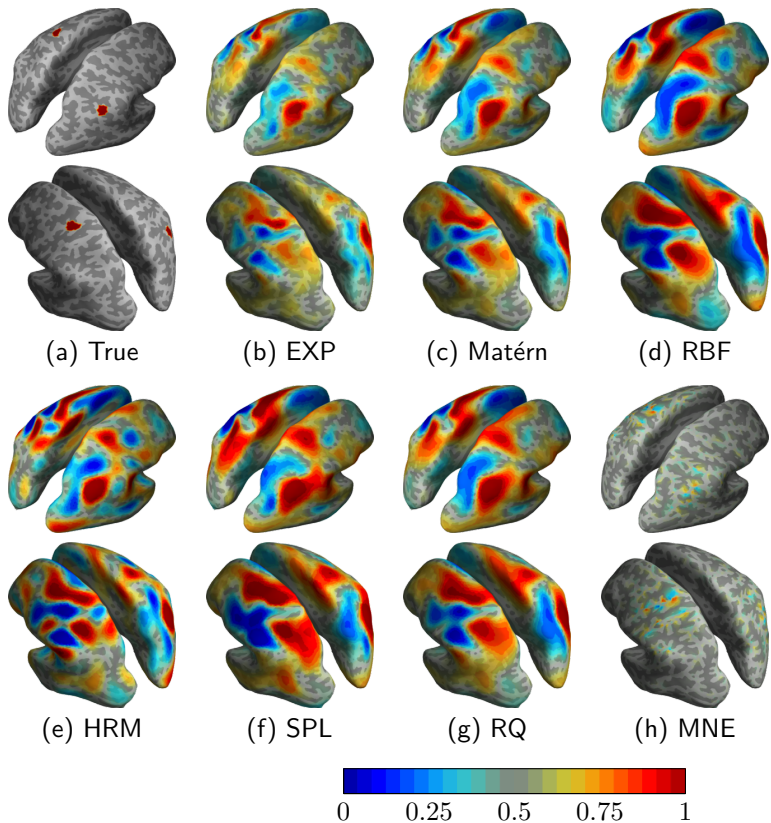


  \tikzsetnextfilename{figure-02}

  \centering\sffamily
  \begin{tikzpicture}

    \def\methodsa{prob-sim1-true,prob-sim1-gpkron-exp,prob-sim1-gpkron-matern32,prob-sim1-gpkron-sexp}
    \def\methodsb{prob-sim1-gpkron-harmony,prob-sim1-gpkron-spl,prob-sim1-gpkron-rq,prob-sim1-mne}
    \def\namesa{{(a)~True},{(b)~EXP},{(c)~Mat\'ern},{(d)~RBF}}
    \def\namesb{{(e)~HRM},{(f)~SPL},{(g)~RQ},{(h)~MNE}}
    \def\xspacing{2.6}

    \foreach[count=\i] \x in \methodsa {
      \node at ({\xspacing*\i},{0}) {\includegraphics[width=2.2cm]{fig/fig-02/\x-a.png}};
      \node at ({\xspacing*\i},{-2.2}) {\includegraphics[width=2.2cm]{fig/fig-02/\x-b.png}};
    }

    \foreach[count=\i] \x in \methodsb {
      \node at ({\xspacing*\i},{-5}) {\includegraphics[width=2.2cm]{fig/fig-02/\x-a.png}};
      \node at ({\xspacing*\i},{-7.2}) {\includegraphics[width=2.2cm]{fig/fig-02/\x-b.png}};
    }

    \foreach[count=\i] \x in \namesa {
      \node at ({\xspacing*\i},{-3.6}) {\x};
    }

    \foreach[count=\i] \x in \namesb {
      \node at ({\xspacing*\i},{-8.6}) {\x};
    }

    \node at (6.05,-9.5) {%
      \begin{axis}[
        hide axis,
        scale only axis,
        height=0cm,
        width=0cm,
        colormap={custom}{[1pt]
          rgb255(0pt)=(0, 0, 153);
          rgb255(1pt)=(0, 0, 204);
          rgb255(2pt)=(0, 0, 255);
          rgb255(3pt)=(0, 51, 255);
          rgb255(4pt)=(0, 102, 255);
          rgb255(5pt)=(7, 152, 248);
          rgb255(6pt)=(34, 184, 221);
          rgb255(7pt)=(60, 195, 195);
          rgb255(8pt)=(103, 168, 152);
          rgb255(9pt)=(125, 141, 130);
          rgb255(10pt)=(130, 141, 125);
          rgb255(11pt)=(152, 168, 103);
          rgb255(12pt)=(195, 195, 60);
          rgb255(13pt)=(221, 184, 34);
          rgb255(14pt)=(248, 152, 7);
          rgb255(15pt)=(255, 102, 0);
          rgb255(16pt)=(255, 51, 0);
          rgb255(17pt)=(255, 0, 0);
          rgb255(18pt)=(204, 0, 0);
          rgb255(19pt)=(153, 0, 0);
        },
        colorbar sampled, samples=20,
        point meta min=0,
        point meta max=1,
        colorbar horizontal,
        colorbar style={
            width=4.5cm,
            height=.33cm,
            xtick={0,.25,.5,0.75,1},
            xtick style={draw=none}
        }]
        \addplot [draw=none] coordinates {(0,0)};
      \end{axis}%
    };

  \end{tikzpicture}
  \caption{Probabilistic posterior summaries of the source estimates. The probability of a source being positive is indicated by the color. This probability equals one if all the marginal probability density is located strictly at the positive real axis and zero in the opposite case. Warm colors indicate positive sources, cold colors negative sources.}
  \label{fig:sim-spatial-prob}
\end{figure*}

Since we developed a Bayesian framework, we can also quantify the uncertainty concerning estimates of the source activations. Figure~\ref{fig:sim-spatial-prob} shows the probability that source being positive, computed using the Gaussian marginal posterior~\eqref{eq:posterior}. Simulations show that taking into account the predictive uncertainty by summarizing $p(\mat{J}_{j,i} \ge 0 \mid \mat{B})$, can convey quite a different message compared to visualizing only the mean $\mathbb{E}[\mat{J}_{j,i}]$.

\begin{figure*}[!ht]

  \tikzsetnextfilename{figure-03}

  \centering\sffamily
  \begin{tikzpicture}

    \def\scales{0,0.01,0.1,0.5}
    \def\spatial{00S,01S,10S,50S}
    \def\temporal{00T,01T,10T,50T}
    \def\xspacing{3}
    \def\yspacing{-2.2}

    \foreach[count=\i] \x in \spatial {
      \foreach[count=\j] \y  in \temporal {
        \node at ({\xspacing*\i},{\yspacing*\j}) {\includegraphics[width=2.5cm]{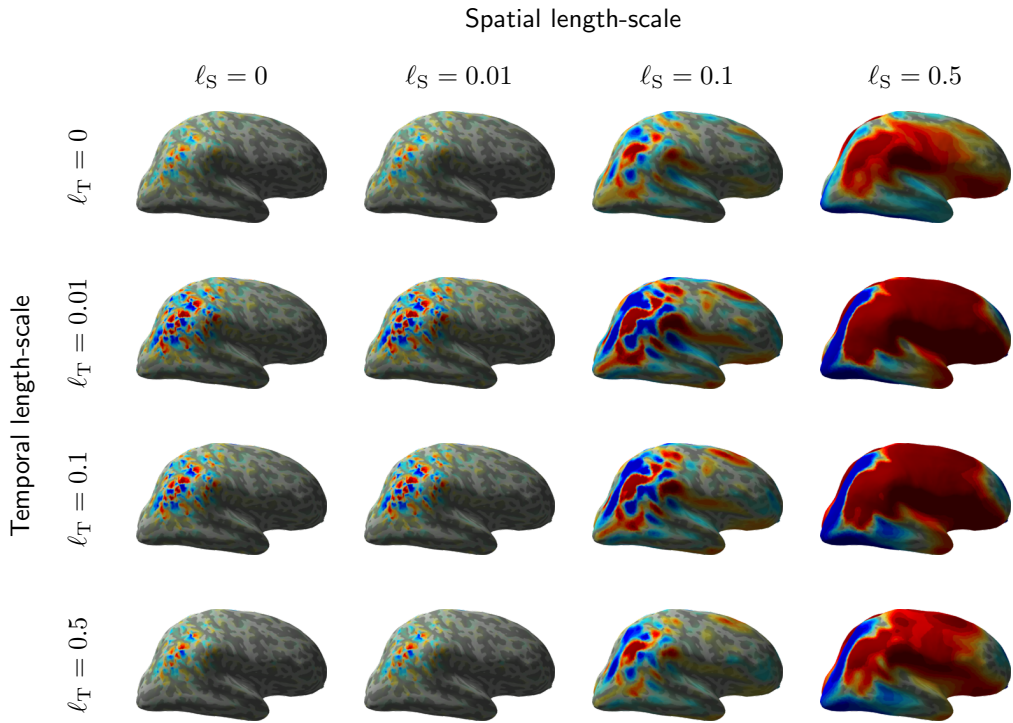}};
      }
    }

    \foreach[count=\i] \x in \scales {
      \node at ({3*\i},-1) {$\ell_\mathrm{S} = \x$};
      \node [rotate=90] at (1,{-2.2*\i}) {$\ell_\mathrm{T} = \x$};
    }

    \node [rotate=90] at (0.25,-5.5) {Temporal length-scale};
    \node at (7.5,-0.25) {Spatial length-scale};

  \end{tikzpicture}
  \caption{Probabilistic posterior summaries in case of increasing spatial and temporal length-scales with the spatio-temporal kernel. Left--right: increasing spatial length-scale ($0, 0.01, 0.1, .5$), top--down: increasing temporal length-scale ($0, 0.01, 0.1, .5$). The upper left corner thus corresponds to MNE. Color coding same as in Figure~\ref{fig:sim-spatial-prob}.}
  \label{fig:sim-length-scale}
\end{figure*}

Given the strong correspondences between different spatially regularised priors, we restrict ourselves to the exponential covariance function in the following. In order to develop an intuition about the influence of the spatial and temporal length scale parameters for the spatio-temporal covariance function~\eqref{eq:spatiotemporal}, Figure~\ref{fig:sim-length-scale} quantifies the uncertainty when varying the length scales.

\subsection{Empirical validation}
In order to empirically validate our approach, as described before, we make use of the data from~\citep{Kauramaki+Jaaskelainen+Hanninen+Auranen+Nummenmaa+Lampinen+Sams:2012}. Our aim is to demonstrate the effect of GP regularization when solving the MEG inverse problem.

We consider a spatially and temporally independent baseline (MNE), a purely temporal model (spatially independent), a purely spatial model (temporally independent), and a spatio-temporal model. We used the exponential covariance function both for the temporal and spatial model. 
We fixed the spatial length-scale parameter to $\ell=0.1$ (recall that we use distances on a unit sphere) and the temporal length scale to $\ell=50$~ms. For each method, subject, and condition we optimized the magnitude hyperparameter $\gamma$ independently by minimising the negative log marginal likelihood.

\begin{figure*}[!t]
  \centering
  \includegraphics[width=\textwidth]{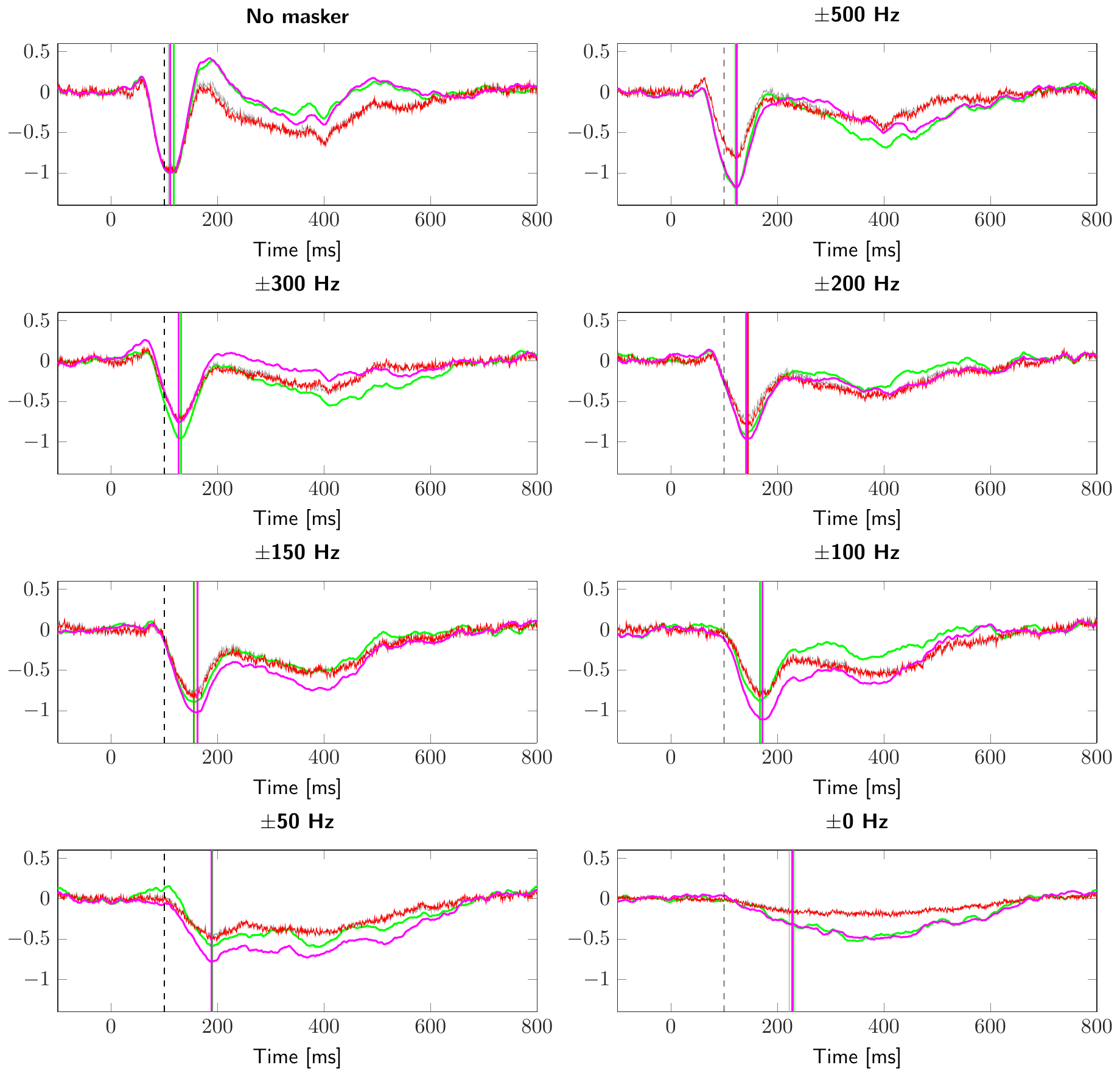}
  \caption{Response suppression due to masking. Grand average ($N = 10$) source waveforms for each stimulus type and condition for the source on the right hemisphere with strongest activation. The MNE reconstruction (grey lines), temporal GP reconstruction (green), spatial GP reconstruction (red), spatio-temporal GP reconstruction (magenta) show similar structure. Peak latency is indicated by vertical lines. The source waveforms illustrate the gradual suppression of response amplitudes and increase in latency with narrower notches. Enhancement of the sustained response is evident, especially in the GP reconstruction.}
  \label{fig:experiment-time-series-norm}
\end{figure*}

\begin{figure*}[!t]
  \centering
  \includegraphics[width=\textwidth]{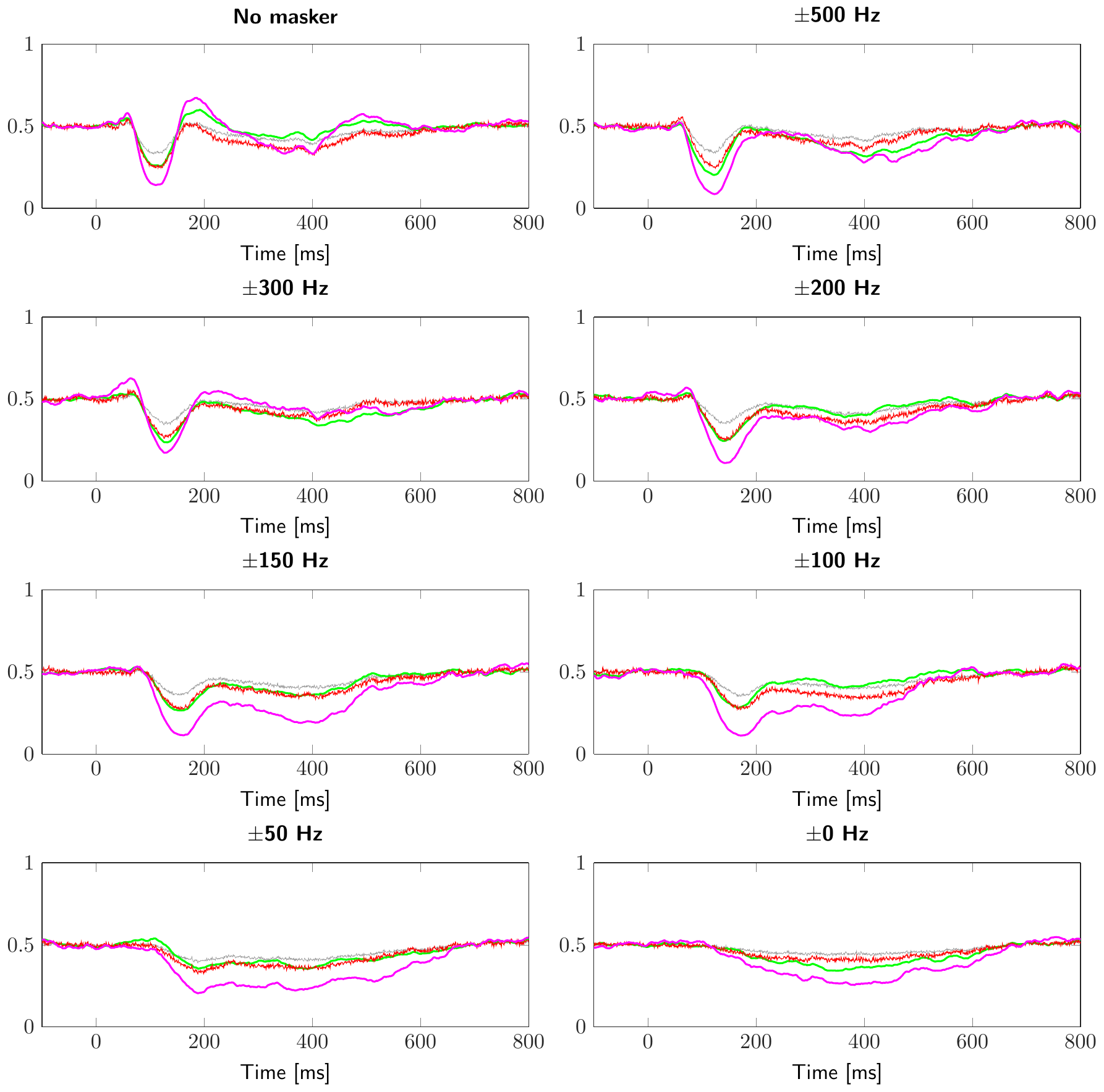}
  \caption{Probabilistic posterior summaries of the temporal activation in the averaged source estimates ($N = 10$) for the source on the right hemisphere with strongest activation. The probability of the source being positive is indicated by the plot. This probability equals one if all the marginal probability density is located strictly at the positive real axis and zero in the opposite case. Color coding as in Figure~\ref{fig:experiment-time-series-norm}.}
  \label{fig:experiment-prob-norm}
\end{figure*}

We study what the source-level ERPs look like when we attend normal vs.\ deviant stimuli in the auditory condition.
Figure~\ref{fig:experiment-time-series-norm} shows the results for response suppression due to masking. Results for the four different methods have been scaled by their M100 peak strength under the `No masker' condition. 
The time-series is shown for the source with the strongest contribution on the right hemisphere.
We observe a gradual suppression of response amplitudes and increase in latency with narrower notches. Slightly different source waveforms are obtained for the different models. Importantly, the spatial and spatio-temporal models show more pronounced enhancement of the sustained response compared to the independent and temporal models. Figure~\ref{fig:experiment-prob-norm} shows the corresponding posterior estimates of observing non-zero source amplitudes. The spatio-temporal model clearly shows the most marked deviations.

\begin{figure*}[!t]

  \tikzsetnextfilename{figure-06}

  \centering\sffamily
  \begin{tikzpicture}

    \def\experiments{sil,500,300,200,150,100,50,0}
    \def\methods{{mne},{gpkron-t},{gpkron-s},{gpkron-st}}
    \def\experimentnames{{No masker},{$\pm500$ Hz},{$\pm300$ Hz},{$\pm200$ Hz},{$\pm150$ Hz},{$\pm100$ Hz},{$\pm50$ Hz},{$\pm0$ Hz}}
    \def\methodnames{{MNE\\(independent)},{Temporal},{Spatial},{Spatio-\\Temporal}}
    \def\xspacing{1.4}
    \def\yspacing{-1.4}

    \foreach[count=\i] \x in \experiments {
      \foreach[count=\j] \y in \methods {
        \node at ({\xspacing*\i},{\yspacing*\j}) {\includegraphics[width=1.3cm]{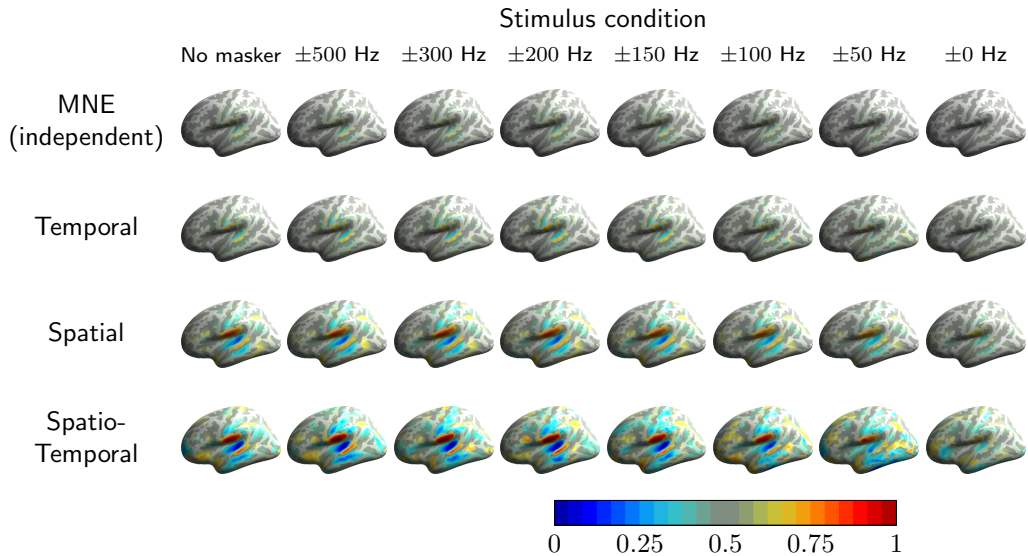}};
      }
    }

    \foreach[count=\i] \x in \experimentnames {
      \node at ({\xspacing*\i},-0.5) {\footnotesize \x};
    }

    \foreach[count=\j] \y in \methodnames {
      \node[align=center] at (-0.5,{\yspacing*\j}) {\y};
    }

    \node at (6.3,0) {Stimulus condition};

    \node at (5.8,-6.8) {%
      \begin{axis}[
        hide axis,
        scale only axis,
        height=0cm,
        width=0cm,
        colormap={custom}{[1pt]
          rgb255(0pt)=(0, 0, 153);
          rgb255(1pt)=(0, 0, 204);
          rgb255(2pt)=(0, 0, 255);
          rgb255(3pt)=(0, 51, 255);
          rgb255(4pt)=(0, 102, 255);
          rgb255(5pt)=(7, 152, 248);
          rgb255(6pt)=(34, 184, 221);
          rgb255(7pt)=(60, 195, 195);
          rgb255(8pt)=(103, 168, 152);
          rgb255(9pt)=(125, 141, 130);
          rgb255(10pt)=(130, 141, 125);
          rgb255(11pt)=(152, 168, 103);
          rgb255(12pt)=(195, 195, 60);
          rgb255(13pt)=(221, 184, 34);
          rgb255(14pt)=(248, 152, 7);
          rgb255(15pt)=(255, 102, 0);
          rgb255(16pt)=(255, 51, 0);
          rgb255(17pt)=(255, 0, 0);
          rgb255(18pt)=(204, 0, 0);
          rgb255(19pt)=(153, 0, 0);
        },
        colorbar sampled, samples=20,
        point meta min=0,
        point meta max=1,
        colorbar horizontal,
        colorbar style={
            width=4.5cm,
            height=.33cm,
            xtick={0,.25,.5,0.75,1},
            xtick style={draw=none}
        }]
        \addplot [draw=none] coordinates {(0,0)};
      \end{axis}%
    };

  \end{tikzpicture}
  \caption{Probabilistic posterior summaries of the spatial activation in the averaged source estimates ($N=10$) at the M100 peak time for each stimulus condition and method. The probability of deviating from zero is indicated with the color. This probability equals one if all the marginal probability density is located strictly at the positive real axis and zero in the opposite case. Warm colors indicate positive sources, cold colors negative sources. The peak time is different for each stimulus condition. Color coding same as in Figure~\ref{fig:sim-spatial-prob}.}
    \label{fig:spatial-activation}
\end{figure*}

The probabilistic posterior summaries of the spatial structure of the activation at the estimated M100 peak corresponding to Figure~\ref{fig:experiment-time-series-norm} can be seen in Figure~\ref{fig:spatial-activation}.

\section{Discussion}
\begin{figure*}[!t]
  \centering
  \includegraphics[width=\textwidth]{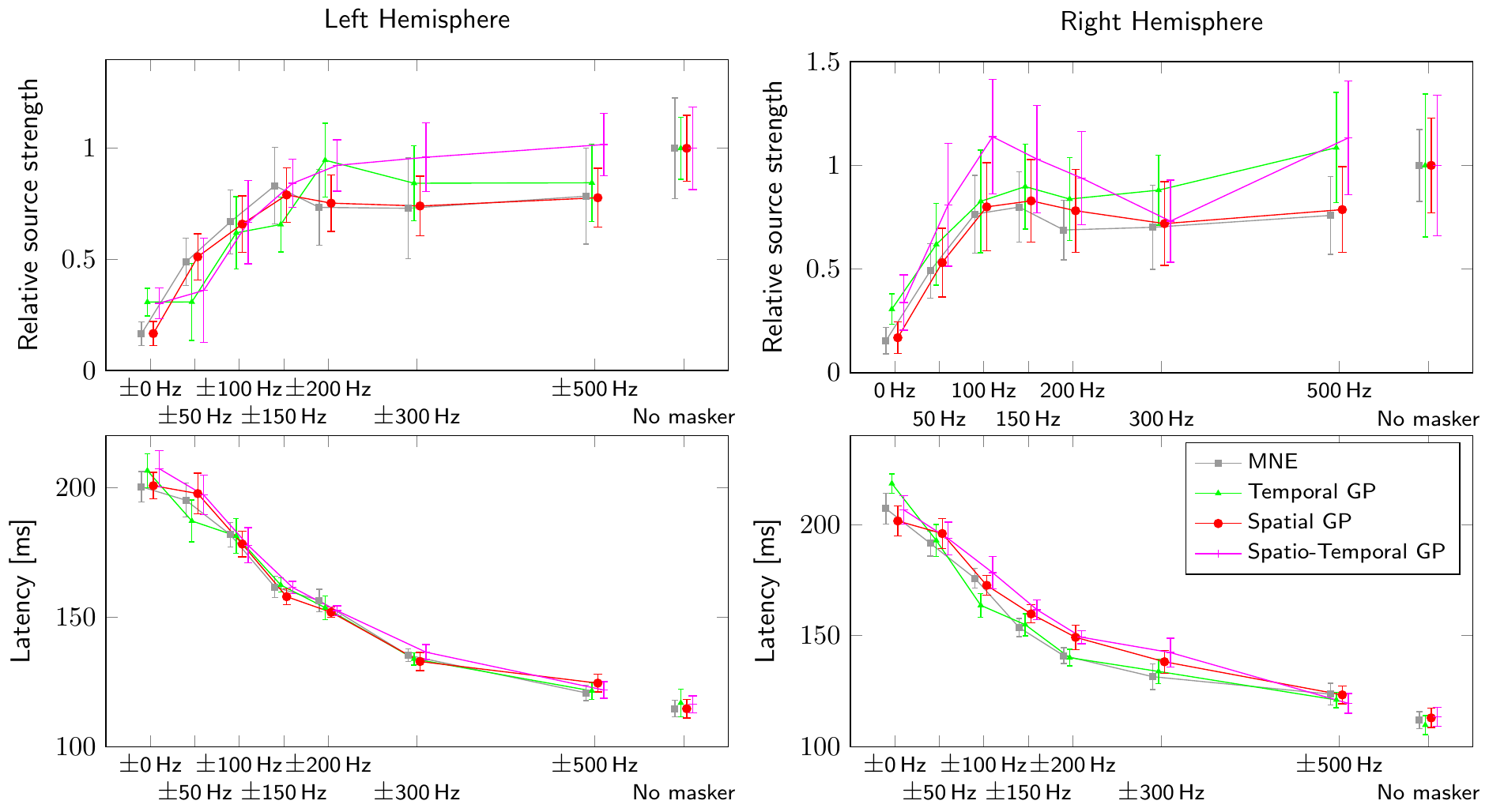}
  \caption{The M100 source strengths ($\pm$SEM) and peak times modulated by masker type (grand average, $N=10$). The source strengths are shown relative to the strength for no masking (far right).}
  \label{fig:M100-strength-time}
\end{figure*}

Here, we compared traditional MNE and a number of spatial covariance selection methods from literature to our spatio-temporal GP framework; specifically how these affect the spread and reliability of the inverse solution. We show that our method is especially good in enhancing weak responses with low SNR, by comparing the results of MNE to regularized solutions with an empirical dataset where SNR was gradually reduced~\citep{Kauramaki+Jaaskelainen+Hanninen+Auranen+Nummenmaa+Lampinen+Sams:2012}. The regularized solutions retain the basic results of peak latency and amplitude modulations (Figures~\ref{fig:experiment-time-series-norm} and \ref{fig:M100-strength-time}) compared to the solution obtained using traditional MNE. Further, these solutions reliably place the activity estimate at a neuroanatomically plausible location, even at low-SNR conditions (Fig.~\ref{fig:spatial-activation}) where MNE estimates have much more uncertainty in auditory areas. The GP-regularized solutions yield spatially smooth estimates, as shown in Figure~\ref{fig:sim-spatial}, and can reliably separate distinct sources with low uncertainty. 

The spatio-temporally regularized inverse solution offers a wealth of applications for neuroscience studies employing MEG, where the interests are towards experiments using more complex, continuously changing natural stimuli, and mapping spatiotemporal specificity of brain areas with ever-higher resolution. One potential benefit of this method is in tracking fast dynamic changes of cortical activity in humans, where EEG and MEG still excel over alternative methods~\citep{Hari+Parkkonen:2015}. Increasing the robustness of inverse solutions can ultimately reduce the number of trials required to obtain a reasonable-SNR averaged evoked response, which allows tracking of small dynamic changes of cortical reactivity during for instance top-down modulation by employing subaveraging of responses throughout the course of an experiment. Further, the possibility to use weak stimulus intensities and low contrasts resulting in low-amplitude responses has a number of implications in mapping human sensory areas with good resolution. For instance, in the auditory domain, the tonotopic organization of human auditory cortex has recently been successfully explored in a high-resolution fMRI experiment with perceptually very weak stimuli~\citep{Langers+Dijk:2012}, where the use of high-intensity stimuli was deemed as a worse option since then the neurons exhibit more non-specific response patterns. Similar response properties apply to macroscopic MEG signals as well, as the underlying neural tuning patterns are the same. Here, our spatio-temporally regularized solution resulted in auditory-cortex estimates with non-zero uncertainty estimates even with the lowest-SNR stimulus from the empirical data (Fig.~\ref{fig:experiment-prob-norm} and \ref{fig:spatial-activation}, stimulus conditions 150~Hz down to 0~Hz, where the latter was set to 50\% detection threshold and subjects were barely able to detect any target sound at all). Importantly, despite poor SNR, the peak estimates with lowest uncertainty were always in the same auditory areas when using the spatio-temporal GP framework (Fig.~\ref{fig:spatial-activation}). This is as expected, as SNR reduction done by introducing a continuous masker sound does not change for the tone-evoked M100 response source location estimated by dipole modeling \citep{Kauramaki+Jaaskelainen+Hanninen+Auranen+Nummenmaa+Lampinen+Sams:2012}. Still, this location invariance could not be verified for these data using simple dipole modeling at lowest SNR conditions without manually setting dipole location constraints. GP solutions, however, provide excellent results in a completely data-driven manner, even for these empirical data. The added robustness using our method can also be useful with special groups used for EEG and MEG studies, such as patients and children, where the evoked-response SNR often cannot be increased by extending the study length and the number of obtained trials.

While obtaining solutions from a few sources is often preferred, we have not considered here how well our GP framework performs with more complex neural response patterns. Here, for instance, the empirical data are collected for auditory stimuli only for a simple stimulus, so the likely contribution is from bilateral auditory areas (only right-hemisphere data shown here). However, when aiming for more complex and natural stimuli, the underlying neural responses can often overlap both spatially and temporally, that is, even for a simple stimulus the response can build up in adjacent areas and with unique temporal dynamics. In these cases, the spatio-temporal prior might hinder the multiple and spatially close source separation with real empirical data. How the regularized solutions fare for more dynamic and complex response patterns remains to be tested in future studies.

Figure~\ref{fig:M100-strength-time} summarizes our empirical findings. It shows M100 sources strengths and peak times modulated by masker type. Note that as we successively consider the independent, temporal, spatial and spatio-temporal models, SEM reduces and curves appear to become more consistent. This further strengthens our belief that spatio-temporal regularization is important for obtaining reliable source estimates.

\section*{Acknowledgements}
This work was supported by the Academy of Finland grant numbers 266940 and 273475, the Netherlands Organization for Scientific Research (NWO) grant numbers 612.001.211 and 639.072.513, and the Finnish Cultural Foundation. We acknowledge the computational resources provided by the Aalto Science-IT project.

\appendix

\section{Proof of Theorem \ref{the:prediction}}
\label{appendixA}

\begin{proof}[Proof of the expression in Equation~\eqref{eq:mean}]
	The following derivation is analogous to the GP formulation for grids presented by \citet{Saatchi:2012} except for the additional linear transformation via the lead field matrix $\mat{G}$ in our setting. 
  As stated in Equation~\eqref{eq:prediction1}, the expression for the mean is given by $\vectb{\mu} = \mathbb{E}[f(\vect{x}_*,t_*)]=\vect{k}_*\T \mat{H}\T 
    (\mat{H} \mat{K} \mat{H}\T 
     + \mat{I})^{-1} \vec{\mat{B}}$. Using the equivalence $\mat{H} = (\mat{I} \kron \mat{G})$ and space-time separability $\mat{K} = \mat{K}_\mathrm{t} \kron \mat{K}_\mathrm{x}$, the mean can be written as
\begin{equation}
  \vectb{\mu} = 
    (\vect{k}_{\mathrm{t}_*}\T \kron \vect{k}_{\mathrm{x}_*}\T)
    (\mat{I} \kron \mat{G})\T
    [(\mat{I} \kron \mat{G})(\mat{K}_\mathrm{t} \kron \mat{K}_\mathrm{x})(\mat{I} \kron \mat{G})\T + \mat{I}\kron\mat{I}]^{-1} 
    \vec{\mat{B}}.
\end{equation}
The transpose is distributive over the Kronecker product,
\begin{equation}
  \vectb{\mu} = 
    (\vect{k}_{\mathrm{t}_*}\T \kron \vect{k}_{\mathrm{x}_*}\T)
    (\mat{I} \kron \mat{G}\T)
    [(\mat{I} \kron \mat{G})(\mat{K}_\mathrm{t} \kron \mat{K}_\mathrm{x})(\mat{I} \kron \mat{G}\T) + \mat{I}\kron\mat{I}]^{-1} 
    \vec{\mat{B}}.
\end{equation}
Using the mixed-product property of the Kronecker product
\begin{equation}
  \vectb{\mu} = 
    (\vect{k}_{\mathrm{t}_*}\T \kron \vect{k}_{\mathrm{x}_*}\T\mat{G}\T)
    (\mat{K}_\mathrm{t} \kron \mat{G}\mat{K}_\mathrm{x}\mat{G}\T 
     + \mat{I}\kron\mat{I})^{-1} \vec{\mat{B}},
\end{equation}
and the eigendecompositions $\mat{K}_\mathrm{t} = \mat{V}_\mathrm{t} \matb{\Lambda}^\mathrm{t} \mat{V}_\mathrm{t}\T$ and $\mat{G}\mat{K}_\mathrm{x}\mat{G}_\mathrm{t}\T = \mat{V}_\mathrm{x} \matb{\Lambda}^\mathrm{x} \mat{V}_\mathrm{x}\T$,
\begin{equation}
  \vectb{\mu} = 
    (\vect{k}_{\mathrm{t}_*}\T\mat{V}_\mathrm{t} \kron 
     \vect{k}_{\mathrm{x}_*}\T\mat{G}\T\mat{V}_\mathrm{x}) 
    \mat{\Lambda}^{-1} (\mat{V}_\mathrm{t}\T \kron \mat{V}_\mathrm{x}\T)\vec{\mat{B}},
\end{equation}
where $\matb{\Lambda} = \matb{\Lambda}^\mathrm{t} \kron \matb{\Lambda}^\mathrm{x} + \mat{I}$. This is a diagonal (non-singular) matrix.

Vectorization combined with a multiplying Kronecker product has the convenient property that if $\mat{A}$ and $\mat{B}$ are nonsingular, $(\mat{A} \kron \mat{B})\vec{\mat{C}} = \vec{\mat{B}\mat{C}\mat{A}\T}$ \citep[see, \eg,][]{Horn+Johnson:2012}. Applying this property once gives us
\begin{equation}
  \vectb{\mu} = 
    (\vect{k}_{\mathrm{t}_*}\T\mat{V}_\mathrm{t} \kron 
     \vect{k}_{\mathrm{x}_*}\T\mat{G}\T\mat{V}_\mathrm{x}) 
    (\mat{\Lambda}^{-1} \vec{\mat{V}_\mathrm{x}\T\mat{B}\mat{V}_\mathrm{t}}).
\end{equation}
The second parentheses contains just a product of a diagonal matrix times a long vector. In order to tidy up the notation, we define a matrix $\matb{\Pi}$ such that $\vec{\matb{\Pi}} = \diag{\matb{\Lambda}^{-1}}$. Element-wise this is $\matb{\Pi}_{j,i} = 1/(\lambda^\mathrm{x}_j \lambda^\mathrm{t}_i + 1)$, where $j=1,2,\ldots, n_\mathrm{n}$ and $i=1,2,\ldots,n_\mathrm{t}$. 
\begin{equation}
  \vectb{\mu} = 
    (\vect{k}_{\mathrm{t}_*}\T\mat{V}_\mathrm{t} \kron 
     \vect{k}_{\mathrm{x}_*}\T\mat{G}\T\mat{V}_\mathrm{x}) 
    \vec{\matb{\Pi} \hadamard \mat{V}_\mathrm{x}\T\mat{B}\mat{V}_\mathrm{t}},
\end{equation}
where `$\hadamard$' denotes the element-wise Hadamard product $\mat{C} = \mat{A} \hadamard \mat{B}$, such that $\mat{C}_{i,j} = \mat{A}_{i,j}\mat{B}_{i,j}$. We apply the product-vectorization property once more to obtain
\begin{equation}
  \vectb{\mu} = 
    \vec{\vect{k}_{\mathrm{x}_*}\T\mat{G}\T\mat{V}_\mathrm{x} 
    (\matb{\Pi} \hadamard \mat{V}_\mathrm{x}\T\mat{B}\mat{V}_\mathrm{t})
    \mat{V}_\mathrm{t}\T \vect{k}_{\mathrm{t}_*}}.
\end{equation}
If $\vect{k}_{\mathrm{x}_*}$ and $\vect{k}_{\mathrm{t}_*}$ are vectors, we can drop the vectorization, and we obtain the result in Theorem~\ref{the:prediction}.%
\end{proof}

\begin{proof}[Proof of the expression in Equation~\eqref{eq:variance}]
From Equation~\eqref{eq:prediction2}, we focus on the latter part, denoted by $\rho = \vect{k}_*\T \mat{H}\T 
    (\mat{H} \mat{K} \mat{H}\T 
     + \mat{I})^{-1} \mat{H} \vect{k}_*$ and write
\begin{align}
    \rho &= (\vect{k}_{\mathrm{t}_*}\T \kron \vect{k}_{\mathrm{x}_*}\T)
    (\mat{I} \kron \mat{G})\T
    [(\mat{I} \kron \mat{G})(\mat{K}_\mathrm{t} \kron \mat{K}_\mathrm{x})(\mat{I} \kron \mat{G})\T + \mat{I}\kron\mat{I}]^{-1}     (\mat{I} \kron \mat{G})
    (\vect{k}_{\mathrm{t}_*} \kron \vect{k}_{\mathrm{x}_*}),
\end{align}
where we apply the distributivity of the transpose in the Kronecker product and the mixed-product property
\begin{equation}
   \rho =(\vect{k}_{\mathrm{t}_*}\T \kron \vect{k}_{\mathrm{x}_*}\T\mat{G}\T)
    [\mat{K}_\mathrm{t} \kron \mat{G}\mat{K}_\mathrm{x}\mat{G}\T 
     + \mat{I}\kron\mat{I}]^{-1} 
    (\vect{k}_{\mathrm{t}_*} \kron \mat{G}\vect{k}_{\mathrm{x}_*}).
\end{equation}
Again, we consider the eigendecompositions $\mat{K}_\mathrm{t} = \mat{V}_\mathrm{t} \matb{\Lambda}^\mathrm{t} \mat{V}_\mathrm{t}\T$ and $\mat{G}\mat{K}_\mathrm{x}\mat{G}_\mathrm{t}\T = \mat{V}_\mathrm{x} \matb{\Lambda}^\mathrm{x} \mat{V}_\mathrm{x}\T$ and get
\begin{equation}
   \rho =(\vect{k}_{\mathrm{t}_*}\T\mat{V}_\mathrm{t} \kron \vect{k}_{\mathrm{x}_*}\T\mat{G}\T\mat{V}_\mathrm{x})
    \matb{\Lambda}^{-1} 
    (\mat{V}_\mathrm{t}\T\vect{k}_{\mathrm{t}_*} \kron \mat{V}_\mathrm{x}\T\mat{G}\vect{k}_{\mathrm{x}_*}).
\end{equation}
This is an inner product, and regrouping the the terms gives us
\begin{equation}
   \rho = [(\vect{k}_{\mathrm{x}_*}\T\mat{G}\T\mat{V}_\mathrm{x}) \hadamard
     (\vect{k}_{\mathrm{x}_*}\T\mat{G}\T\mat{V}_\mathrm{x})]
    \matb{\Pi}
    [(\mat{V}_\mathrm{t}\T \vect{k}_{\mathrm{t}_*}) \hadamard
     (\mat{V}_\mathrm{t}\T \vect{k}_{\mathrm{t}_*})].
\end{equation}
That is, we get the result in Theorem~\ref{the:prediction}.%
\end{proof}

\bibliographystyle{model5-names}
\bibliography{bibliography}

\end{document}